\documentclass[
 reprint,twocolumn,
 amsmath,amssymb,
 aps,nofootinbib,superscriptaddress,
]{revtex4-2}

\usepackage{amsthm}
\usepackage{graphicx}
\usepackage{color}
\usepackage{dcolumn}
\usepackage{bm}
\usepackage{mathtools}      
\usepackage{mathrsfs}       
\usepackage{braket}
\usepackage{float}
\usepackage[table]{xcolor}

\newcommand{\cF}{\ensuremath{\mathcal{F}}}

\newcommand{\cN}{\ensuremath{\mathcal{N}}}

\newcommand{\cL}{\ensuremath{\mathcal{L}}}
\newcommand{\cW}{\ensuremath{\mathcal{W}}}

\newcommand{\cX}{\ensuremath{\mathcal{X}}}

\newcommand{\cS}{\ensuremath{\mathcal{S}}}

\newcommand{\bx}{\ensuremath{\boldsymbol{x}}}
\newcommand{\by}{\ensuremath{\boldsymbol{y}}}

\newcommand{\bn}{\ensuremath{\boldsymbol{n}}}

\newcommand{\bX}{\ensuremath{\boldsymbol{X}}}
\newcommand{\bY}{\ensuremath{\boldsymbol{Y}}}

\newcommand{\bU}{\ensuremath{\boldsymbol{U}}}
\newcommand{\bI}{\ensuremath{\boldsymbol{I}}}
\newcommand{\bW}{\ensuremath{\boldsymbol{W}}}

\newcommand{\bb}{\ensuremath{\boldsymbol{b}}}

\newcommand{\bu}{\ensuremath{\boldsymbol{u}}}
\newcommand{\bw}{\ensuremath{\boldsymbol{w}}}

\newcommand{\bbR}{\ensuremath{\mathbb{R}}}

\DeclareMathOperator{\hsig}{\hat{\sigma}}

\newcommand{\hc}{\hat{\ensuremath{c}}}
\newcommand{\hcd}{\hat{\ensuremath{c}}^{\dagger}}
\newcommand{\ha}{\hat{\ensuremath{a}}}
\newcommand{\had}{\hat{\ensuremath{a}}^{\dagger}}
\newcommand{\hx}{\hat{\ensuremath{x}}}
\newcommand{\hxd}{\hat{\ensuremath{x}}^{\dagger}}

\DeclareMathOperator{\tr}{\textup{Tr}}

\DeclareMathAlphabet{\mymathbb}{U}{BOONDOX-ds}{m}{n}

\usepackage{hyperref}
\hypersetup{
    colorlinks = true,
    linkcolor = {blue},
    citecolor = {blue},
    urlcolor = {blue},
}

\begin{document}
\title{Quantum\text{--}Classical Hybrid Information Processing \\via a Single Quantum System}

\author{Quoc Hoan Tran}
\email{k09tranhoan@gmail.com}
\affiliation{Next Generation Artificial Intelligence Research Center (AI Center), Graduate School of Information Science and Technology, The University of Tokyo, Japan}

\author{Sanjib Ghosh}
\email{sanjibghosh@baqis.ac.cn}
\affiliation{
	Beijing Academy of Quantum Information Sciences, Beijing, China
}

\author{Kohei Nakajima}
\email{k-nakajima@isi.imi.i.u-tokyo.ac.jp}
\affiliation{Next Generation Artificial Intelligence Research Center (AI Center), Graduate School of Information Science and Technology, The University of Tokyo, Japan}

\date{\today}

\begin{abstract}
Current technologies in quantum-based communications bring a new integration of quantum data with classical data for hybrid processing.
However, the frameworks of these technologies are restricted to a single classical or quantum task, which limits their flexibility in near-term applications.
We propose a quantum reservoir processor to 
harness quantum dynamics in computational tasks requiring both classical and quantum inputs.
This analog processor comprises a network of quantum dots in which quantum data is incident to the network and classical data is encoded via a coherent field exciting the network.
We perform a multitasking application of quantum tomography and nonlinear equalization of classical channels. 
Interestingly, the tomography can be performed in a closed-loop manner via the feedback control of classical data.
Therefore, if the classical input comes from a dynamical system, embedding this system in a closed loop enables hybrid processing even if access to the external classical input is interrupted.
Finally, we demonstrate preparing quantum depolarizing channels as a novel quantum machine learning technique for quantum data processing.

\end{abstract}

\pacs{Valid PACS appear here}

\maketitle
\section*{Introduction}
\begin{figure*}
		\includegraphics[width=14cm]{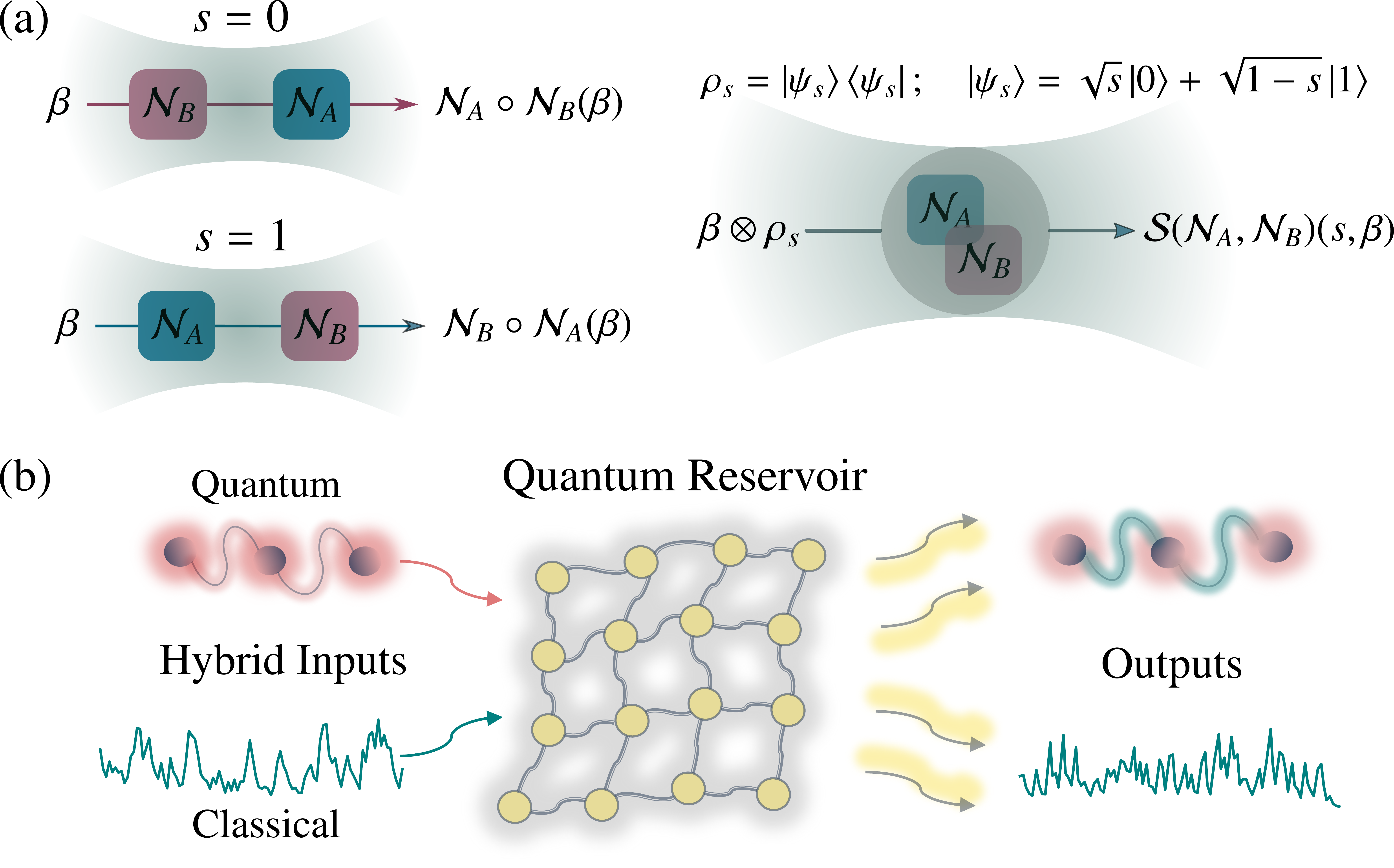}
		\protect\caption{A quantum reservoir processor for quantum-classical hybrid data processing. (a) An example of a quantum device with hybrid inputs. Here, we consider a quantum switch that includes two quantum channels $\cN_A$ and $\cN_B$ and an independent switch state $\rho_s$ controlled by a classical signal $s$. 
		This quantum switch can be considered a function of the hybrid input $(s, \beta)$.
		Given a quantum state $\beta$, the quantum switch produces an output $\cN_A\circ \cN_B(\beta)$ if  $\rho_s = \ket{0}\bra{0}$ (s=0) and $\cN_B\circ\cN_A(\beta)$ if $\rho_s = \ket{1}\bra{1}$ (s=1). 
		When $\rho_s$ is in a superposition of $\ket{0}$ and $\ket{1}$, such as $\rho_s=\ket{\psi_s}\bra{\psi_s}$ with $\psi_s = \sqrt{s}\ket{0}+\sqrt{1-s}\ket{1}$ ($0\leq s \leq 1$), the output becomes a quantum superposition of two alternative orders $\cN_A\circ\cN_B(\beta)$ and $\cN_B\circ\cN_A(\beta)$. 
		(b) Our quantum reservoir (QR) is a network of quantum dots that can receive both quantum and classical data as input. 
		Quantum inputs are incident via optical fields, and classical inputs are encoded in experimental control fields.
		The appropriate readout after a time evolution on QR can provide a high-dimensional transformation for both classical and quantum inputs, which can be used in learning tasks.
		\label{fig:hybrid:overview}}
\end{figure*}

Recent advances in machine learning (ML) and quantum computing have revolutionized the methodology of processing complex and large-scale data.
While merging these fields, classical or quantum systems can generate a massive amount of time series data, such as sensing data or quantum states that flow through multiple quantum channels in a network of quantum devices~\cite{kimble:2008:quinternet,simon:2017:nat:quinternet,wehner:2018:science:quinternet}.
This context leads to the requirement of a novel learning paradigm to process these data efficiently,
such as the easy manipulation used in training and deployment, while maintaining rich representation capability.
Currently, algorithms are being designed on specific homogeneous data, such as quantum-native or classical-native data.
However, most quantum devices rely on classical controls~\cite{viola:1999:prl:open,dong:2010:control}, such as temperature or signals from electronic controllers~\cite{vandijk:2019:control,rist:2020:microwave}.
The outputs of these devices are not simply derived from quantum channels and are also considered a function of classical controls and quantum input.
A representative example is a quantum switch with classical control, which simulates the indefinite causal order between two operations~\cite{chiribella:2012:qswitch,procopio:2015:qswitch:exp,rubino:2017:qswitch:exp,goswami:2018:qswitch:exp,wei:2019:qswtich:exp} [Fig.~\ref{fig:hybrid:overview}(a)].
Therefore, the research on hybrid quantum and classical data processing can lead to broader and near-term applicability for quantum devices. For example, we can use the same resource to learn the tomography of devices receiving both classical and quantum data without doing it separately for each control setting.

Contrary to ML models such as artificial recurrent neural networks on a digital computer, a physical system with rich dynamics can be a good candidate for a learning system within the framework of physical reservoir computing (PRC)~\cite{nakajima:2021:RCbook,nakajima:2020:physical}. In PRC, the input is fed into a dynamical system called a reservoir to create nonlinear dynamics of input data via sufficiently complex and high-dimensional trajectories~\cite{jaeger:2001:echo,jaeger:2001:short,maass:2002:reservoir,lukoeviius:2009:reservoir,nakajima:2021:RCbook}.
A readout, which outputs a linear combination of the accessible observables in the reservoir, is the only part that needs to be trained without interfering with the reservoir's internal parameters.
Accordingly, the success and efficiency of PRC rely on good physical realizations of the reservoir, which has attracted considerable interest from diverse research fields~\cite{nakajima:2020:physical}.
The seminal work~\cite{fujii:2017:qrc} uses a disordered ensemble quantum dynamics system as a quantum reservoir (QR) to process classical data,
with the possibility of having a large number of degrees of freedom.
QRs have been developed in various platforms, such as nuclear magnetic resonance (NMR) systems~\cite{fujii:2017:qrc,nakajima:2019:qrc,tran:2020:higherorder}, superconducting quantum processors~\cite{chen:2020:temporal,suzuki:2021:natural}, fermions and bosonic models~\cite{ghosh:2019:quantum,ghosh:2020:reconstruct,ghosh:2019:neuromorphic,khan:2021:qrc:boson},  quantum harmonic oscillators~\cite{nokkala:2020:gaussian,gerasimos:2021:prx:measurement},
arrays of Rydberg atoms~\cite{bravo:2021:rydberg}, and photonic quantum memristors~\cite{spagnolo:2022:natphotonic}.
Several studies have focused on the processing of data in the form of quantum states~\cite{ghosh:2019:quantum,ghosh:2019:neuromorphic,ghosh:2020:reconstruct,ghosh:2021:quantum:adv,tran:2021:temporal,nokkala:2021:online}, which provide certain advantages over classical ML methods. 
However, a QR is yet to be treated as a homogeneous data-driven model because it lacks the ability to deal with hybrid forms of quantum-classical data.
Therefore, an unified architecture for hybrid quantum-classical processing is required from theoretical and applied perspectives. 

In this study, we establish a framework that considers a QR as an analog processor to process hybrid quantum-classical data.
Inspired by Refs.~\cite{ghosh:2019:quantum,ghosh:2020:reconstruct}, our QR is a network of quantum dots with random inter-site couplings.
Classical inputs are encoded in classical controls, such as coherent pumps in the network,
and quantum inputs are incident to the QR in the form of optical fields. For temporal processing, each quantum input interacts with the QR for a short duration before being replaced by another input.
The time evolution of the interactions provides a high-dimensional nonlinear mapping of the input via the correlations in the QR, which can be extracted by classical or quantum readouts on accessible nodes.
This enables us to learn the function of input sequence, leading to diverse applications in classical and quantum data processing.

\section*{Results}

\textbf{Quantum\text{--}Classical Hybrid Information Processing via a Quantum Reservoir. }
When we describe a quantum device processing quantum data in a realistic scenario, we must incorporate classical control into the model.
In this case, a quantum device is in fact a function of quantum input $\beta$ and classical control $u$ as $\cF(u, \beta)$, where we consider the scalar $u$ for ease of explanation.
For a device processing the sequence of hybrid inputs $(u_1, \beta_1)$, $(u_2, \beta_2)$, \ldots, we can describe it using the temporal map $\cF(\{(u_l, \beta_l)\})$ of input history~\cite{tran:2021:temporal}.
Our target is to develop a trainable framework to emulate $\cF$.

The proposed framework contains three main parts: an input part containing input modes to receive the data, a QR processor to interact with inputs in a quantum evolution, and a readout for further processing [Fig.~\ref{fig:hybrid:overview}(b)]. We consider the QR processor as a two-dimensional lattice of $N$ quantum dots, represented by the Hamiltonian
\begin{align}\label{eqn:hamiltonian}
    \hat{H} = &\sum_i E_i\hcd_i\hc_i + \sum_{\langle i,j \rangle}h_{ij}\left( \hcd_i\hc_j +  \hcd_j\hc_i\right)\nonumber\\
   & + \sum_iQ_i\hcd_i\hcd_i\hc_i\hc_i + P(t)\sum_i\left( \hcd_i + \hc_i \right),
\end{align}
where $\hc_i$, $E_i$, $h_{ij}$, $Q_i$, and $P(t)$ are the field operators, onsite energies, hopping amplitudes between the nearest neighbor sites, nonlinearity strengths, and uniform time-dependent coherent field strengths, respectively. 
$P(t)$ can be used to encode the classical input $u(t)$ as 
$P(t) = P + Wu(t)$, where $P$ and $W$ are the constant coefficient and input scaling, respectively.

The dynamics of the combined quantum state $\rho$ of the QR as well as the input modes can be described by the quantum master equation (we use the unit where Plank constant $\hbar=1$).
\begin{align}
    \dot{\rho} = -i[\hat{H}, \rho] + \gamma\sum_j\cL(\hc_j)\rho + \Omega(t - t_{\text{init}})\hat{A}\rho,
\end{align}
where $\Omega(t)=1$ for $t\geq 0$ and $0$ otherwise.
Here, $\hat{A}\rho=\sum_k \dfrac{\gamma_k}{\gamma}\cL(\ha_k)\rho + \sum_{k,j}W_{jk}^{\text{in}} \left( \left[\ha_k\rho, \hcd_j \right] + \left[\hc_j, \rho\had_k \right] \right)$ represents the cascade coupling between the input modes $\ha_k$ and the QR~\cite{gardiner:1993:prl:driving}. 
The Lindblad superoperator $\cL(\hat{x})$ is defined for any arbitrary operator $\hat{x}$ by $\cL(\hx)\rho=\hx\rho\hxd - \dfrac{1}{2}\left( \hxd\hx\rho + \rho\hxd\hx\right)$.

\begin{figure*}
		\includegraphics[width=17cm]{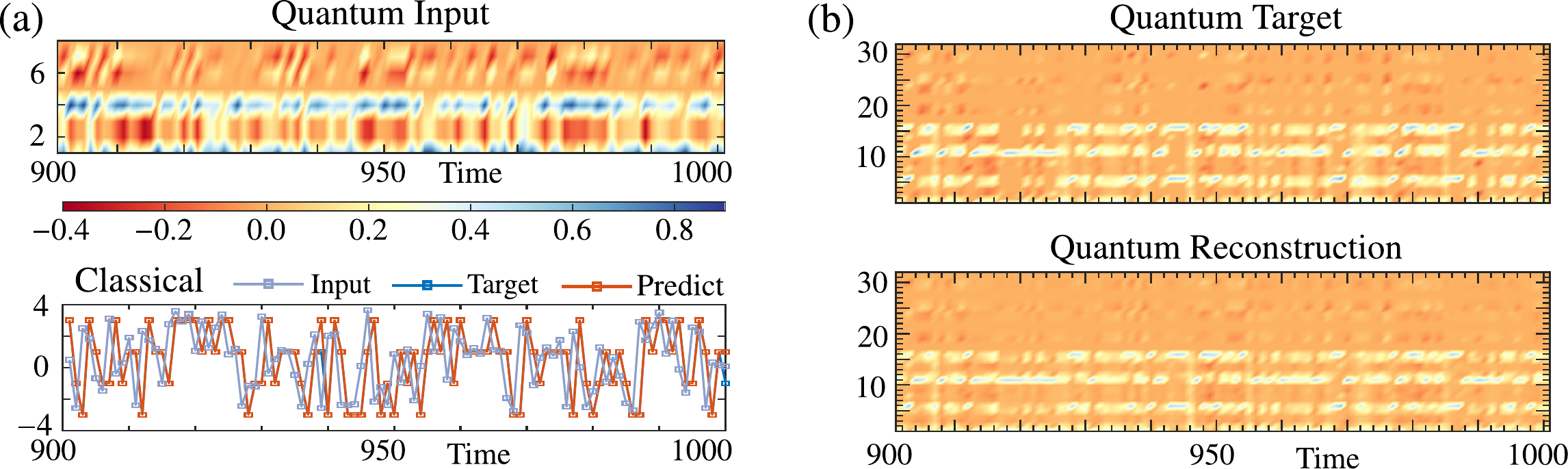}
		\protect\caption{Demonstration of the quantum tomography task and the classical channel equalizer task. (a) A random sequence of one-qubit quantum inputs (upper panel) and a result for the channel equalizer task (bottom panel) in the evaluation phase. Each quantum state is represented as a  real vector by stacking the real and imaginary parts of the density matrix. (b) The target and reconstructed tomography with $N=3$ reservoir sites, $P/\gamma=0.1$, $W/\gamma=1.0$, and the measurement multiplexity $V=8$.
		\label{fig:switch:eqn}}
\end{figure*}

\begin{figure*}
		\includegraphics[width=17cm]{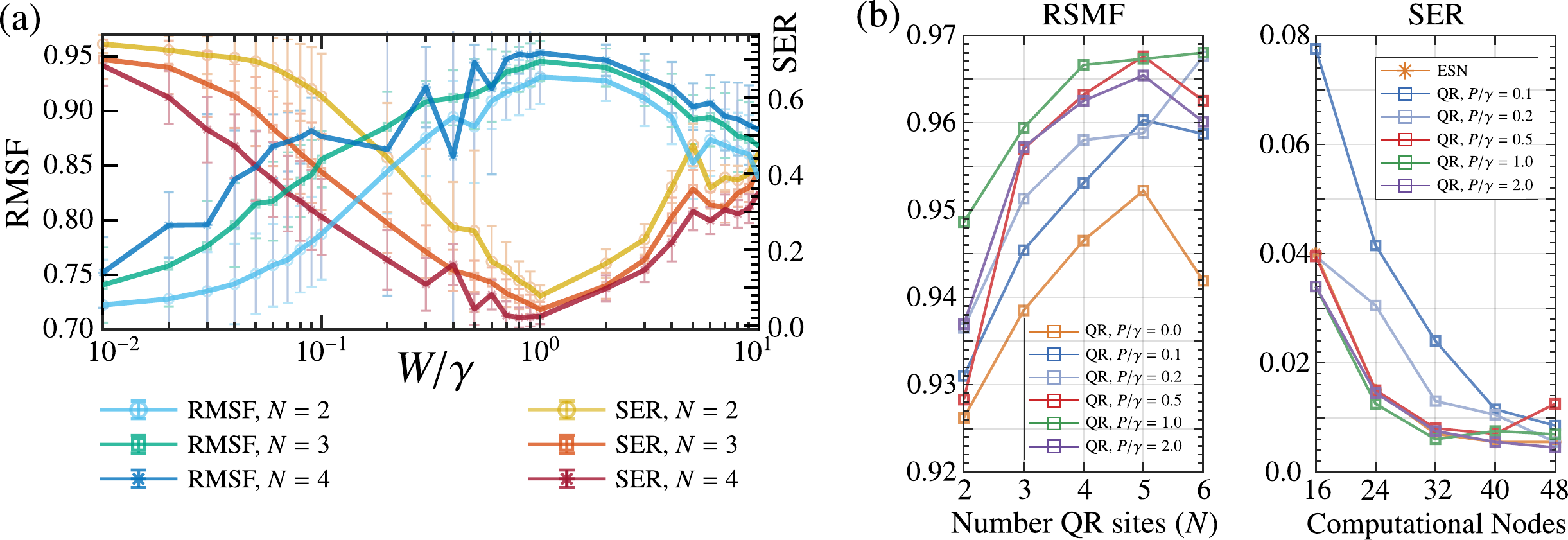}
		\protect\caption{Performance in the quantum tomography and classical channel equalizer tasks. (a) The average root mean square of fidelities (RMSF) and the average symbol error rate (SER) with shaded error bars over 10 trials.
		(b) (Left) The average RSMF in the tomography task when we increase the number $N$ of reservoir sites in the QR.  
		(Right) Comparison between the average SER in the Echo State Network (ESN) and in our QR for the same number of computational nodes. 
		In (b), we set the input scaling as $W/\gamma=1.0$ and the measurement multiplexity as $V=8$ for numerical experiments; therefore, the QR with number of computational nodes $16, 24, 32, 40, 48,$ corresponds to $N=2, 3, 4, 5, 6$ sites in the reservoir.
		\label{fig:switch:perform}}
\end{figure*}

We explain quantum-classical hybrid processing using the proposed platform.
First, the QR is excited only with the uniform $P$ for $0\leq t< t_{\text{init}}$ and no incident quantum inputs. We choose $t_{\textup{init}}$ such that the QR at time $t_{\text{init}}$ reaches a steady state.
This setting ensures the echo state property~\cite{jaeger:2001:echo} for the reproducible computation, where the response to the same input sequence is independent of the QR's initial state. 
Then, the quantum input $\beta$ (described by the input modes $\ha_k$) is incident to the reservoir, and the classical input $u(t)=u$ is activated at the same time.
At time $t_1 = t_{\textup{init}} + \tau$ for time interval $\tau$, an appropriate and practical readout from the QR is performed for nontrivial transformations of input data
(see Supplementary Information for detailed settings of $h_{ij}$, $\gamma_k$, $W_{jk}^{\text{in}}$, $\tau$, and $t_{\textup{init}}$).
We consider two readout schemes: a linear combination of measurement results on the accessible observables (classical readout) and the other with a linear combination of quantum modes (quantum readout). The former is associated with a measurement process, while the latter has been considered in a quantum neuromorphic platform for quantum state preparation~\cite{ghosh:2019:neuromorphic}. 

For a non-temporal processing task, we repeat the above procedure for every hybrid data instance $(u, \beta)$.
For a temporal processing task, at $t_l = t_{\textup{init}} + (l-1)\tau$ ($l=1,2,\ldots$), the classical input is switched to $u(t)=u_l$, and the quantum state $\beta_l$ replaces the partial state in the input modes. Since the input information is transferred into the QR during the interaction, this scheme enables the memory ability, which is required in temporal processing tasks.

In the classical readout, measuring the expectation values of the occupation numbers $n_j=\langle \hcd_j\hc_j\rangle$ can extract the information from the QR to reconstruct $\cF$.
A representative application is quantum tomography, which reconstructs the density matrix output of $\cF$ via the linear regression model: $W^{\textup{out}}\bn + \bb \approx \bY_{\cF}$~\cite{ghosh:2020:reconstruct,tran:2021:temporal}. 
Here, $\bn=(n_1,\ldots,n_K)^\top$ is the $K$-dimensional reservoir state for readout, $\bY_{\cF}$ is the real vector form to stack the real and imaginary elements of $\cF$, 
and $W^{\textup{out}}$ and $\bb$ are the weight and bias parameters to be optimized via the training (see Methods).
In the classical readout, multitasking is possible since the training cost is minimal for independent training with different $W^{\textup{out}}$ for different tasks.
If the measurement is performed after an interaction time $\tau$ for the current input and right before the next input, the dimensionality $K$ is equal to the number of quantum dots $N$.
One can increase this dimensionality by performing measurements at different timings in the interval $\tau$, which is known as the temporal multiplexing technique.
Between two inputs, we perform measurements at equal interval $\tau/V$, forming the dimensionality $K=NV$.
Here, $V$ is called the measurement multiplexity. 
Another technique to increase the dimensionality $K$ is spatial multiplexing~\cite{nakajima:2019:qrc}, where  readout reservoir states in different QRs are combined to learn the target.

In the quantum readout, the standard toolbox of linear optical elements~\cite{braunstein:2005:toolbox} enables us to generate $M$ quantum output modes 
$
    \hat{C}_m = \sum_j o_{mj}\hc_j
$ with complex coefficients $o_{mj}$.
The output modes must satisfy the commutation relations $[\hat{C}_m, \hat{C}_n^{\dagger}] = \delta_{mn}$, which impose the condition $\sum_j o_{mj}o^{*}_{nj}=\delta_{mn}$.
Since the target is the quantum state, the training process is not as simple as the one used for linear regression on the accessible observables in the classical readout.
Consider the separation of non-adjustable and adjustable parameters in PRC, we assume that the parameters of Hamiltonian in Eq.~\eqref{eqn:hamiltonian} are random and not trainable. Instead, we train interaction ($W^{\textup{in}}_{jk}$) and readout ($\{o_{mj}\}$) coefficients such that the quantum state described via $\{\hat{C}_m\}$ becomes the same as the output of $\cF$ (see Methods).

\textbf{Quantum Tomography and Channel Equalizer.} We present an application of QR to hybrid tasks in which quantum tomography and noise-free reconstruction of classical data are performed simultaneously.
Consider a temporal map $\cF\{(s_l, \beta_l)\}$ where $\{s_l\}$  and $\{\beta_l\}$ are the sequences of classical controls and quantum inputs, respectively.
We assume that the output state  $\cF_l = \cF\{(s_l, \beta_l)\}$ is accessible at $l=1,\ldots,L$ for training. The tomography task learns the relation between $\cF_l$ and $\{(s_l, \beta_l)\}$ for $l \leq L$ and reconstructs $\cF_l$ with $l>L$.
Obviously, the QR cannot learn this hybrid task without the information contained in $\{s_l\}$.
Therefore, we further assume that the classical control data are also accessible, although only in a distorted form of a nonlinear transformation $s_l\to u_l$.
Since multitasking is feasible in the classical readout, we can also reconstruct $\{s_l\}$ from $\{u_l\}$.

In the following example, we consider $\cF$ as a quantum switch with classical control [Fig.~\ref{fig:hybrid:overview}(a)].
Technically, a quantum switch includes two quantum channels $\cN_A$ and $\cN_B$ representing the operations by Alice and Bob, respectively, and an independent switch state $\rho_s$.
Signal communication between Alice and Bob is only restricted to a partial order.
However, the quantum switch can send the information under the indefinite causal order of quantum channels~\cite{chiribella:2012:qswitch,procopio:2015:qswitch:exp,rubino:2017:qswitch:exp,goswami:2018:qswitch:exp,wei:2019:qswtich:exp}.
Given a state $\beta$ on which these channels act, the quantum switch produces an output $\cN_A\circ \cN_B(\beta)$ if  $\rho_s = \ket{0}\bra{0}$ and $\cN_B\circ\cN_A(\beta)$ if $\rho_s = \ket{1}\bra{1}$.
When the switch state is in a superposition of $\ket{0}$ and $\ket{1}$, such as $\rho_s=\ket{\psi_s}\bra{\psi_s}$ with $\psi_s = \sqrt{s}\ket{0}+\sqrt{1-s}\ket{1}$ ($0\leq s \leq 1$), the output becomes a quantum superposition of two alternative orders $\cN_A\circ\cN_B(\beta)$ and $\cN_B\circ\cN_A(\beta)$.
Here, the quantum switch $\cS(\cN_A, \cN_B)$ can be considered a function of hybrid input $(s, \beta)$.

We use our QR to mimic the behavior of the quantum switch applied to the input sequence. 
Given a delay $d$, we demonstrate that the QR with current inputs $\beta_l$ and $u_l$ can utilize memory effects to reconstruct $\sigma_l=\cS(\cN_A, \cN_B)(s_{l-d}, \beta_{l-d})$ and $s_{l-d}$.
We consider $\cN_A, \cN_B$ as two depolarizing quantum channels and the reconstruction of $\{s_{l-d}\}$ from $\{u_l\}$ as the nonlinear channel equalization task (see Methods).
Here, $\{\beta_l\}$ is an i.i.d. sequence of one-qubit density matrices,  and  $\{s_l\}$ is an i.i.d. discrete sequence of symbols, which are selected from $\{-3, -1, 1, 3\}$ with equal probability. The switch state at each $l$ is $\rho_{(3+s_l)/6}$, and the distorted input $\{u_l\}$ is transformed from $\{s_l\}$ via both linear and nonlinear channels~\cite{jaeger:2004:harness} (see Methods).
If $d\geq 1$, it requires a QR with the nonlinear effect and a memory of both quantum and classical inputs.

The QR's output is divided into two parts: the tomography result $\bY_l$ in the real vector form and the equalized result $y_l$.
$\bY_l$ is then transformed in the density matrix form $\hat{\sigma}_l$ with the consideration of a projection to obtain a positive semidefinite matrix (see Methods).
$y_l$ is converted back into a nearest symbol $\hat{s}_l \in \{-3, -1, 1, 3\}$.
The training is performed at $l=1,\ldots,L$ ($L=800$), and the tomography performance is evaluated via the root mean square of fidelities (RMSF)
\begin{align}
    \textup{RMSF}=\sqrt{(1/T)\sum_{l=L+1}^{l=L+T}F^2(\sigma_l, \hat{\sigma}_l)},
\end{align}
where $T=200$ and $F(\rho,\sigma )=\tr[\sqrt{\sqrt{\sigma}\rho\sqrt{\sigma}}]$.
The equalization performance is evaluated via the symbol error rate (SER)
\begin{align}
\text{SER}=\textup{card}(\{l \mid \hat{s}_l  \neq s_{l-d}\})/T.
\end{align}

Figure~\ref{fig:switch:eqn}(a) illustrates a sequence of one-qubit quantum input in the evaluation phase (upper panel) and a result for the channel equalizer (bottom panel) with delay $d=1$ (see Supplementary Information for results with other $d$ values). 
Here, the predict and target sequences for the reconstruction of classical symbols $\{s_l\}$ are overlapped at almost all time steps.
The density matrix at each time step is represented as a real vector by stacking the real and imaginary parts.
Figure~\ref{fig:switch:eqn}(b) depicts that the quantum target sequence can be reconstructed well. 

We systematically evaluate the performance of the tomography and channel equalizer tasks via the RMSF (left axis) and SER (right axis) in Fig.~\ref{fig:switch:perform}(a) for different $N$ and $W$.
A large value of $W$ compared with $h_{ij}$ and $W_{jk}^{\text{in}}$ leads to non-ergodic behavior in the QR, i.e., a strong and qualitative dependence on the initial state at $t_{\textup{init}}$ (Fig.~S1 in Supplementary Information).
In addition, in Supplementary Information, we further investigate the effects of the classical input in the reconstruction of the quantum input.
With a large $W$, the input state is incident with weak coupling ($|W_{jk}^{\textup{in}}| \ll |P(t)=P+Wu(t)|$) under a strong effect of the classical input to the QR's dynamics, which means that not much information regarding quantum inputs can be retained in the QR.
In contrast, a small $W$ reduces the memory effect in reconstructing the previous classical input.
This explains the existence of a region of $W$ for an optimal performance ($W/\gamma\approx 1.0$).

The left panel of Fig.~\ref{fig:switch:perform}(b) displays the RMSF of the tomography task when we increase the number $N$ of reservoir sites.  
In the right panel of Fig.~\ref{fig:switch:perform}(b), we further compare the performance in the equalization task with the Echo State Network (ESN) in classical reservoir computing under the condition of the same number of computational nodes (see Methods). Here, we set the input scaling as $W/\gamma=1.0$ and use the QR with the measurement multiplexity $V=8$; therefore, the QR containing $16, 24, 32, 40, 48$ computational nodes corresponds to $N=2, 3, 4, 5, 6$ sites in the reservoir. 
We confirm that with appropriate setting of the constant coherent field $P$, we can obtain almost the same  performance with the ESN.

\textbf{Continuous Variable Tomography and Closed Loop.}
We modify the situation in the tomography task where, after the training phase, we were unable to access the information from the classical control $s_l$.
Surprisingly, owing to the advantages of multitasking, our QR can autonomously generate $s_l$ in a closed-loop manner while performing the tomography task with the hybrid input.
In the training phase, $s_l$ is learned in an open loop where we predict the next step $s_{l+1}$ given the input $u_l=s_l$.
After training, the prediction is used as the classical input for the next step, forming a closed-loop control without any external interventions.
This model-free prediction is well established in classical reservoir computing, for example, to predict low-dimensional chaotic systems~\cite{jaeger:2004:harness} or large spatiotemporally chaotic systems~\cite{pathark:2018:prl:chaos}.
However, to the best of our knowledge, our demonstration is the first to combine the closed-loop setting with the quantum tomography task, which is only effective in the QR setting.

\begin{figure*}
		\includegraphics[width=17cm]{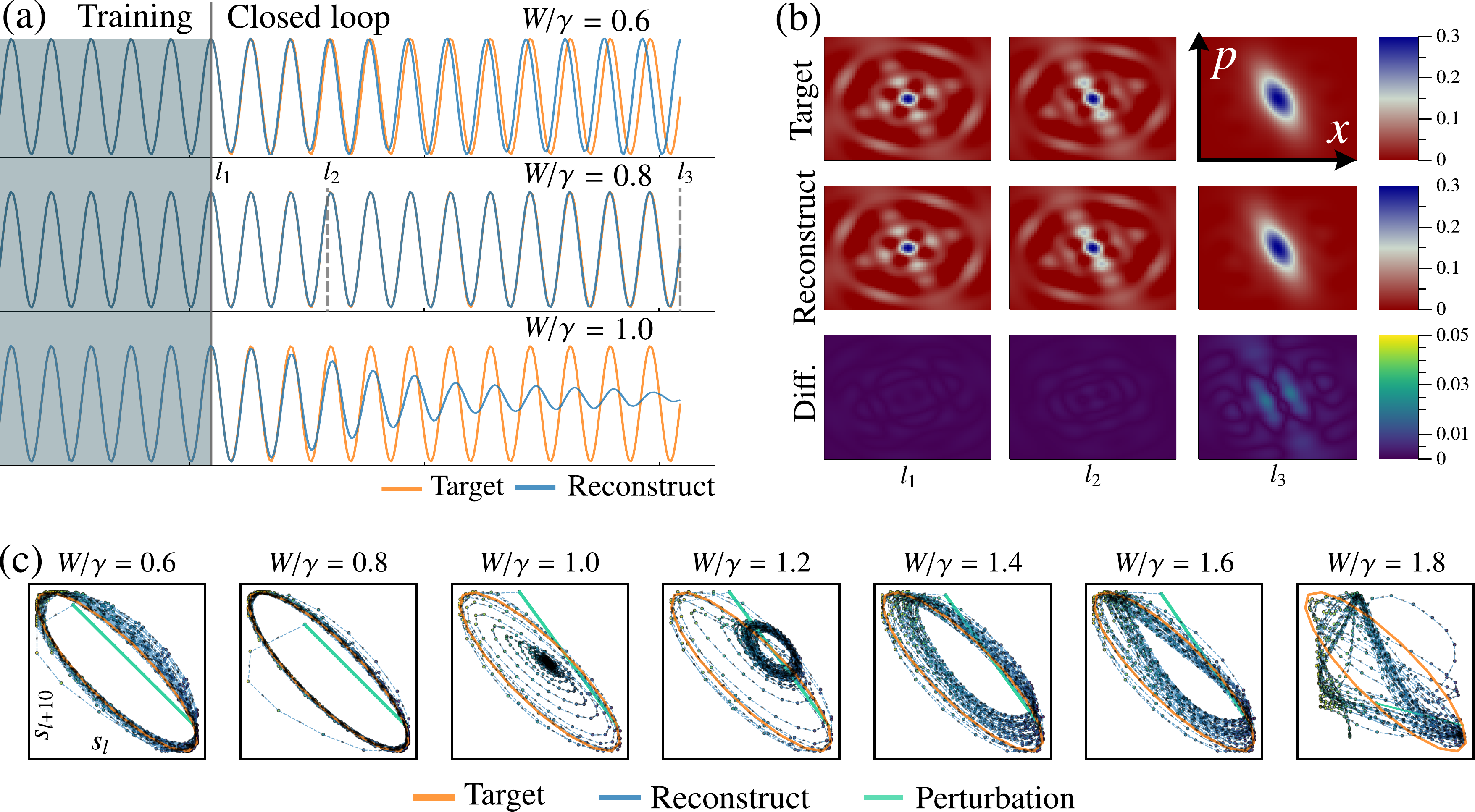}
		\protect\caption{Continuous variable tomography and closed-loop control of periodic classical signals. (a) Closed-loop control of classical signals with $N=3$, $V=10$, $P/\gamma=1.0$, and $W/\gamma=0.6,0.8,$ and $1.0$. (b) Continuous variable tomography at typical time steps in the closed loop with $W/\gamma=0.8$. The last panel displays the absolute difference between the target and reconstructed Wigner functions.
		(c) Stability after adding a small perturbation to the trajectory for different input scaling $W/\gamma$.
		\label{fig:close:amp}}
\end{figure*}

We consider the quantum tomography of continuous variable states. 
The target is to reconstruct the output $\cF_l= \cF\{(s_l,\beta_l)\}$ in the Wigner function form $\cW(\cF_l; x_i, p_j)$ defined on a grid of continuous variables $x_i$ and $p_j$ (see Methods).
We use 300 randomly generated one-mode thermal states $\beta_l$ and the periodic signals $s_l = 0.5 + 0.5\sin(\dfrac{l\pi f}{510})$ in the training phase.
The target $\cF_l$ is created by applying one-mode squeezing operator to $\beta_l$ as
\begin{align}
    \hat{S}(\xi_l)=\exp\left( \xi_l \had\had - \xi_l^{*}\ha\ha \right),
\end{align}
where $\xi_l=s_le^{i\pi/4}$.
In Supplementary Information, we consider another encoding: $\xi_l=0.3e^{i2\pi s_l}$.
Here, we consider the cutoff Fock space dimension (the effective dimension) of these continuous variables states is $D_{\textup{eff}}=9$.

Figure~\ref{fig:close:amp}(a) shows examples of the control signals in the training and closed-loop phase for $f=60$. With $W/\gamma=0.8$ and $N=3$ sites, the control signal is almost reconstructed perfectly for all time steps in the closed-loop phase. 
This QR can efficiently reconstruct the Wigner function even without accessing the control signal [Fig.~\ref{fig:close:amp}(b)].
We further investigate the stability of the closed-loop trajectories plotted in the $(s_l, s_{l+10})$ plane [Fig.~\ref{fig:close:amp}(c)].
The QR presents a stable embedding of sinusoidal classical inputs if the trajectory can return to the target after adding a small perturbation (green line) into a predicted value, suggesting that our system successfully learned the target attractor.
We observe an appropriate setting of input scaling $W$ to obtain stable closed loops ($W/\gamma \approx 0.8)$.
Intriguingly, if we increase $W/\gamma$, for example to $W/\gamma = 1.8$, the closed loop fails to reconstruct the trajectory of the sinusoidal input in the evaluation stage but can produce chaotic-like behavior in the embedding space.
In this case, the generated trajectory is not elliptical as the trajectory of sinusoidal inputs but still robust with respect to a small perturbation.
We also observe the dependency of the performance of closed-loop control and the production of  chaotic-like behavior on time scales $f$ of the control signals, which are investigated in detailed in Supplementary Information.

\textbf{Quantum Readout and Depolarizing Channel.}
Finally, we present an application using the quantum readout scheme to output quantum states. We use the QR to prepare a depolarizing quantum channel $\cF\{(s_l, \beta_l)\} = s_{l}I/D + (1-s_{l})\beta_{l}$, where $\{\beta_l\}$ are randomly generated in a $D$-dimensional Hilbert space and $\{s_l\}$ is a random sequence in $[0, 1]$.

\begin{figure*}
		\includegraphics[width=17cm]{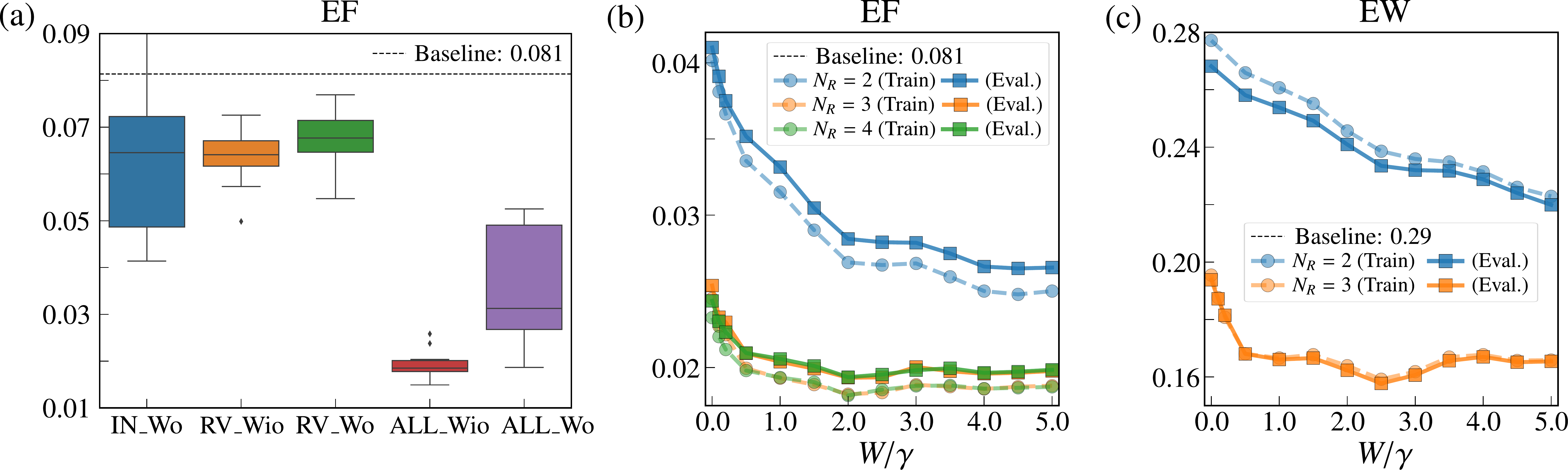}
		\protect\caption{The error in training and evaluating the quantum readout to prepare the depolarizing quantum channel. (a) Combinations of readout nodes and training parameters, where only input modes (IN), only reservoir modes (RV), or both of them (ALL) are considered as $N_R$ readout nodes. 
		The training parameters are readout weights (Wo) or both readout weights and interaction coefficients (Wio). Fidelity error with one-qubit input states (b) and error taken in the Wigner representation with continuous variable states (c) varying with input scaling $W/\gamma$. 
		\label{fig:qtask:depolar1b}}
\end{figure*}

First, we consider a sequence of $200$ one-qubit quantum states for the training and $100$ states for the evaluation. The baseline is computed when we set the output as the same as the input.
We use the Nelder\text{--}Mead simplex algorithm~\cite{lagarias:1998:nd} (see Methods) to minimize 
the fidelity error
\begin{align}\label{eqn:fidelity:error}
    \textup{EF}=\sqrt{\dfrac{1}{L}\sum_{l=1}^{L}[1-F(\sigma_l, \hat{\sigma}_l)]^2},
\end{align}
where $\sigma_l$ and $\hat{\sigma}_l$ are the target and preparing quantum states, respectively.
In Fig.~\ref{fig:qtask:depolar1b}(a), the evaluated fidelity errors with different readout and training configurations are presented for the QR with $N=2$ sites, $P=1.0$, and $W=2.0$. The interquartile range is contained within the box, and the 5th and 95th percentiles are marked by whiskers. The median is the line across the box, and the outliers are located outside the whiskers of each box plot.
Here, IN, RV, and ALL correspond to the setting where only input modes $\ha_k$, only reservoir modes $\hc_j$, or both of them are considered as the readout nodes, respectively. Wo and Wio correspond to the situation where only readout weights or both readout weights and interaction coefficients $W^{\textup{in}}_{jk}$ are considered as the training parameters, respectively.
The result implies that the consideration of both input and reservoir modes as $N_R$ readout modes and both interaction coefficients and readout weights for training leads to the best performance. 
Under this setting, we display the variation in fidelity errors EF with the input scaling  $W/\gamma$ and $N_R$ in  Fig.~\ref{fig:qtask:depolar1b}(b).
Even with a small QR ($N_R=3, 4$) we can prepare the target channel with a relatively low error ($<2\%$), which is significantly better compared with the baseline ($\approx 8\%$).
Furthermore, increasing $W/\gamma$ basically leads to a better performance where more information regarding the classical input is integrated.

Finally, we prepare the depolarizing channel using the input quantum states as random squeezed thermal states in the continuous variable form. We minimize the cost function taken in the Wigner representation as follows:
\begin{align}\label{eqn:wigner:error}
    \textup{EW} = \sqrt{ \dfrac{1}{L}\sum_{l=1}^L\dfrac{\sum_{i,j}[\cW(\sigma_l; x_i, p_j) - \cW(\hat{\sigma}_l; x_i, p_j)]^2}{\sum_{i,j}[\cW(\sigma_l; x_i, p_j) + \cW(\hat{\sigma}_l; x_i, p_j)]^2}}.
\end{align}
Owing to the scale limitation, we only simulate the continuous variable states of the effective dimension $D_{\textup{eff}}=3$, where $D=D^2_{\textup{eff}}=9$.
Figure~\ref{fig:qtask:depolar1b}(c) presents the errors in 50 training and 50 evaluating data varying with $W$. 
We can observe a similar trend in Fig.~\ref{fig:qtask:depolar1b}(b), that is, with sufficient classical information ($W/\gamma\geq 1.0$), the error EW ($\approx 0.16$) with $N_R=3$ readout nodes is significantly lower than the baseline's error ($\approx 0.29$).
This result is still below a considerably good preparation ($\textup{EW}< 0.1$), but it demonstrates that hybrid inputs can be effectively considered for training the quantum readout.

\section*{Discussion}
We proposed a framework for an analog QR processor with hybrid inputs and classical and quantum readouts for learning heterogeneous quantum-classical data.
This aligns well with scenarios where one wishes to model a quantum device to process quantum input but must rely on classical control signals in physical experiments.
Our framework, therefore, has the potential to be physically implemented in quantum network systems where classical control and quantum sources can interact with nonlinear quantum systems to form a quantum channel. 
It can help realize quantum adaptive systems capable of quantum information processing.
These agents can be used to interpret and memorize both classical and quantum signals from their environment and to respond accordingly to the actions of their surroundings~\cite{elliott:2022:quantum:adaptive}.

Processing hybrid quantum and classical data is a promising idea to facilitate future innovative use cases for quantum computers.
This concept aims to leverage the advantages of quantum mechanics in ML with an unconventional computing framework and intriguing applications.
It is not limited to the conventional discussion on practical quantum advantages, such as the ``beating speedup" of quantum to classical ML methods~\cite{schuld:2022:advantage}. 
For example, classical readouts lead to interesting applications of multitasking where quantum data can be processed in a closed loop of the classical control. 
Furthermore, adding this closed-loop mechanism allows us to utilize the unique coherence properties of quantum systems to generate unique classical dynamics.
We consider the quantum readout to avoid the measurement process of preparing the quantum output. However, optimization can be challenging and requires improvement, since we need to simulate or drive the quantum system and evaluate the cost function for a wide range of parameters.

A further enticing discussion would be the case of the correlation between the processing of quantum data and classical data in a hybrid setting of the QR. We can consider a QR to simultaneously process quantum data and classical data as separate tasks. An intriguing research question arises: Can this multitasking mechanism induce positive or negative effects on information processing?
For example, if we repeatedly modify the coherent field strengths of the QR via a classical input with a large magnitude, it can limit the short-term memory properties of quantum data processing (see Supplementary Information).
However, one can also expect positive effects and not only negative ones. There may exist a situation where simultaneously processing different modals of data can actually bring an optimal regime rather than solely solving a single task.
We can start by investigating relations between hybrid input protocols with the dynamics of the QR, such as the classical input may induce the dynamical phase transition in the QR~\cite{pena:2021:qrc:dynamic}.
We can also study how classical and quantum data are processed via the QR's dynamics, such as by decomposing the readout reservoir states in terms of basis polynomials for input history~\cite{dambre:2012:nonlinear,kubota:2021:prr:IPC}.
Along with this research line, one can refer to a recent study demonstrating that quantum noise in real quantum processors can induce the information processing capability when using classical data~\cite{kubota:2022:noise}.

\providecommand{\noopsort}[1]{}\providecommand{\singleletter}[1]{#1}%

\section*{Methods}
\textbf{Reservoir Computing (RC).}
RC is based on the modeling for the transformation of the input sequence in high-dimensional trajectories~\cite{jaeger:2001:echo,jaeger:2001:short,maass:2002:reservoir,lukoeviius:2009:reservoir}.
RC contains three main parts: the input part to store the input sequence, the reservoir, and the readout part.
A reservoir is a dynamical system driven by an input to encode recurrent relations and nonlinear dynamics of the input history.
The readout part with connections to the reservoir can extract reservoir states, which are useful features retained by the reservoir for emulating the target sequence.
In general, we only need to train the readout connections, making RC particularly suitable for physical implementations.

Mathematically, RC is described by the input-driven map $g: \bU \times \cX \to \cX \subset \mathbb{R}^K$,
where $\bU$ and $\cX$ are the input and the reservoir's state space, respectively.
Here, $K$ is considered the dimension of the reservoir's state.
If we feed a discrete-time input sequence $\{\ldots, \bu_{-1}, \bu_0, \bu_1, \ldots\}$ into the reservoir, the readout reservoir state $\bx_l$ is represented by the following recurrent relation:
\begin{align}\label{eqn:driven-map}
\bx_l = g(\bu_l, \bx_{l-1}).
\end{align}

In temporal supervised learning tasks, we are given an input sequence $\{\bu_1,\ldots,\bu_L\}$ and the corresponding target sequence $\hat{\by}=\{\hat{\by}_1,\ldots,\hat{\by}_L\}$, where $\hat{\by}_k \in \bbR^d$ with $d$ is the output dimension.
We consider a parameterized readout map $h_{\bw}: \cX \to \bbR^d$, where the output signal is $\by_l = h_{\bw}(\bx_l)$.
The readout map is often taken as a linear combination of the readout reservoir states as $\by_l = h_{\bw}(\bx_l) = \bw^\top\bx_l$.
Here, $\bw$ is the trainable parameter obtained by minimizing the error between $\by_l$ and $\hat{\by_l}$ over $l=1,\ldots,L$. For example, we often consider the mean square error
\begin{align}
    \textup{MSE} = \dfrac{1}{L}\sum_{l=1}^L \Vert \by_l - \hat{\by}_l \Vert^2_2,
\end{align}
where $\Vert \cdot \Vert_2$ denotes the Euclidean norm between two vectors in $\bbR^d$.
For training, we add a constant bias term $x_{l,K+1}=1$ to the readout reservoir state $\bx_l$
and optimize $\bw$ via the linear regression $\hat{\bY} = \bX\bw$. 
Here, $\hat{\bY}=[\hat{\by}_1\quad\ldots\quad \hat{\by}_L]^\top$ is the $L\times d$ target matrix and
$\bX$ is the $L\times (K+1)$ matrix that combines the readout reservoir states $\bx_1, \bx_2, \ldots, \bx_L$ of the training data.
The optimal value of $\bw$ is obtained via the Ridge regression in the matrix form $\boldsymbol{\hat{w}^\top} = (\bX^\top\bX + \eta\bI)^{-1}\bX^\top\hat{\bY}$, where $\eta$ is a positive constant for the regularization.

\textbf{Echo State Network (ESN). }
ESN is a realization of the input-driven map in RC.
It belongs to the concept of artificial recurrent neural network (RNN), as we have a large network with randomly fixed inner and recurrent connections.
Consider ESN with $N$ computational nodes and a discrete-time input sequence $\{\bu_l\}$, the reservoir state $\bx_l$ at time step $l$ is described by 
\begin{eqnarray}
    \bx_{l+1} = \textbf{tanh}\left(\bW^{\textup{in}}\bu_{l+1} + \bW\bx_l \right),
    \label{eq:esn}\nonumber
\end{eqnarray}
where $\textbf{tanh}(\cdot)$ is the activation function applied on vector $\bx=(x_1, x_2, \ldots, x_K)^\top$ as $\textbf{tanh}(\bx) = (\tanh(x_1), \ldots, \tanh(x_K))^\top$. 
Here, $\bW^{\textup{in}}$ and $\bW$ are the input weight matrix and recurrent weight matrix, respectively.
In the channel equalization task, the input weight matrix $\bW^{\textup{in}}$ is generated from random uniform distribution in $[-1, 1]$.
We also set the connection probability and the spectral radius of the recurrent weight matrix $\bW$ to 0.1 and 0.9, respectively.

\textbf{Learning Quantum Tomography. }
A quantum device can be described by a function of quantum input $\beta$ and classical control $u$ as $\cF(u, \beta)$, where we consider the scalar $u$ for ease of notation.
Given a sequence of hybrid inputs $(u_1, \beta_1)$, $(u_2, \beta_2)$, \ldots and a quantum device with a time-dependent behavior, we can describe it using the temporal map $\cF_l = \cF(\{(u_i, \beta_i)\}_{i=1:l})$ of the current and past inputs~\cite{tran:2021:temporal}.
We assume that we have full tomography for the corresponding output states of $\cF$ in the training, where 
we are given a hybrid input sequence  $\{(u_1, \beta_1),\ldots,(u_L, \beta_L)\}$ and the corresponding target sequence $\hat{\by}=\{\hat{\by}_1,\ldots,\hat{\by}_L\}$.
Here, $\hat{\by}_l$ is the real vector form of $\cF_l$.
If $\cF_l$ is described by the density matrix,  $\hat{\by}_l$ is formed by stacking the real and imaginary elements of $\cF(\beta_l)$.
In the evaluation stage, we are given an input sequence $\{(u_{L+1}, \beta_{L+1}), \ldots, (u_{L+T}, \beta_{L+T})\}$ with the corresponding target $\{\hat{\sigma}_{L+1},\ldots,\hat{\sigma}_{L+T}\}$, where $\hat{\sigma}_i = \cF(u_i, \beta_i)$.
The output sequence is $\{{\by}_{L+1}, \ldots, {\by}_{L+T}\}$, which is rearranged in the matrix form $\{{\sigma^{\prime}}_{L+1}, \ldots, {\sigma^{\prime}}_{L+T}\}$. 
To obtain the final positive semidefinite matrix $\sigma_i$,
we project $\sigma^{\prime}_i$ onto the spectrahedron such that the trace of $\sigma_i$ is equal to 1 and the Frobenius norm of $\sigma_i - \sigma^{\prime}_i$ is minimized~\cite{chen:2011:projection,bantysh:2020:quantum}.

Tomography learning can be performed with other forms of $\cF_l$, for example, in the Wigner function representation of continuous variable states.
Given a density matrix $\sigma$, the continuous variable states associated with $\sigma$ can be described by the Wigner function
\begin{align}
    W\left(\sigma; x_{i}, p_{j}\right)=\int \frac{d y}{2 \pi}\left\langle x_{i}+\frac{y}{2}\left|\sigma\right| x_{i}-\frac{y}{2}\right\rangle e^{-i y p_{j}},
\end{align}
where the integral is evaluated in the whole space and the states $\ket{x_i\pm \dfrac{y}{2}}$ represent continuous position bases. We evaluate Wigner functions on a $61\times 61$ grid of $x_i$ and $p_j$, where we divide the interval $[-3, 3]$ into $60$ equal intervals for the range of $x_i$ and $p_j$.
The target of continuous variable tomography is to reconstruct these Wigner functions, i.e., the real $61\times 61$ dimensional matrices.

\textbf{Quantum Switch. }
In the classical counterpart, a switch is an operation of control that can route a target system through two classical channels $C_A$ and $C_B$ in series following one causal order ($C_A$ then $C_B$) or the other ($C_B$ then $C_A$).
The quantum switch is different in that it induces entirely new quantum trajectories where the order of the two operators is indefinite.
Technically, a quantum switch includes two quantum channels $\cN_A$ and $\cN_B$ to create a new channel $\cS(\cN_A, \cN_B)$, which uses the channels $\cN_A$ and $\cN_B$ in an order that is entangled with an independent switch quantum state $\rho_s$. 
The channel $\cS(\cN_A, \cN_B)$ returns the state $(\cN_A\circ\cN_B(\rho))\otimes \ket{0}\bra{0}$ if $\rho_s = \ket{0}\bra{0}$ and $(\cN_B\circ\cN_A(\rho))\otimes \ket{1}\bra{1}$ if $\rho_s = \ket{1}\bra{1}$.
When $\rho_s$ is in a superposition of $\ket{0}$ and $\ket{1}$, the channel returns a correlated state as a result of $\cN_A$ and $\cN_B$ acting on $\rho$ in a superposition of two alternative orders.

To describe $\cS(\cN_A, \cN_B)$, we denote the Kraus operators of channels $\cN_A$ and $\cN_B$ as $\{K^{(A)}_i\}$ and $\{K^{(B)}_j\}$, respectively, where
$\cN_A = \sum_{i}K^{(A)}_i\rho K^{(1)\dagger}_i$ and $\cN_B = \sum_{j}K^{(B)}_j\rho K^{(2)\dagger}_j$.
The Kraus operators of $\cS(\cN_A, \cN_B)$ are defined as
\begin{align}
    W_{ij} = K^{(A)}_iK^{(B)}_j \otimes \ket{0}\bra{0} + K^{(B)}_jK^{(A)}_i \otimes \ket{1}\bra{1}.
\end{align}
The action of the quantum switch is given by
\begin{align}
    \cS(\cN_A, \cN_B)(\rho \otimes \rho_s) = \sum_{i,j}W_{ij}(\rho \otimes \rho_s)W^{\dagger}_{ij}.
\end{align}
In our study, we consider $\cN_A$ and $\cN_B$ as two depolarizing channels with parameters $q_A$ and $q_B$, which are given by
\begin{align}
    \cN_A(\rho) &= (1-q_A)\rho + q_A \frac{I}{D}  = (1-q_A)\rho + \dfrac{q_A}{D^2}\sum_{i=1}^{D^2}U_i\rho U^{\dagger}_i
    \nonumber\\ &= \dfrac{q_A}{D^2}\sum_{i=0}^{D^2}U_i\rho U^{\dagger}_i,\\
    \cN_B(\rho) &= (1-q_B)\rho + q_B \frac{I}{D} = (1-q_B)\rho + \dfrac{q_B}{D^2}\sum_{j=1}^{D^2}V_j\rho V^{\dagger}_j 
    \nonumber \\ &= \dfrac{q_B}{D^2}\sum_{j=0}^{D^2}V_j\rho V^{\dagger}_j,
\end{align}
where $D\times D$ is the dimension of $\rho$ and $\{U_i\}_{i=1}^{D^2}$,  $\{V_j\}_{j=1}^{D^2}$ are orthonormal bases of the space of $D\times D$ matrices. 
Here, we introduce the notation $U_0 = \dfrac{D\sqrt{1-q_A}}{\sqrt{q_A}} I$ and $V_0 = \dfrac{D\sqrt{1-q_B}}{\sqrt{q_B}} I$.
We define the extension Kraus operators for $\cN_A$ and $\cN_B$ as $K^{(A)}_i = \dfrac{\sqrt{q_A}}{D}U_i$ for $i=0,1,\ldots,D^2$ and $K^{(B)}_j = \dfrac{\sqrt{q_B}}{D}V_j$ for $j=0,1,\ldots,D^2$, respectively.
We can express the Kraus operators of $\cS(\cN_A, \cN_B)$ as
\begin{align}
    W_{ij} = \dfrac{\sqrt{q_Aq_B}}{D^2}\left( U_iV_j \otimes \ket{0}\bra{0} + V_jU_i \otimes \ket{1}\bra{1}\right).
\end{align}

We consider the control state $\rho_s = \ket{\psi_s}\bra{\psi_s}$, where $\psi_s = \sqrt{s}\ket{0} + \sqrt{1-s}\ket{1}$ ($0\leq s \leq 1$). The output of the quantum switch is given by
\begin{widetext}
\begin{align}\label{eqn:qswitch}
    &\cS(\cN_A, \cN_B)(\rho \otimes \rho_s) = {A}^{00}\otimes\ket{0}\bra{0} + {A}^{01}\otimes\ket{0}\bra{1} + {A}^{10}\otimes\ket{1}\bra{0} + {A}^{11}\otimes\ket{1}\bra{1}, 
\end{align}
where
\begin{align}
    {A}^{00} &= s\dfrac{q_Aq_B}{D^4} \sum_{i=0}^{D^2}\sum_{j=0}^{D^2}U_iV_j\rho V_j^{\dagger}U_i^{\dagger} = s\cN_A\cN_B(\rho) \\
    {A}^{01} &= A^{10} = \sqrt{s (1-s)}\dfrac{q_Aq_B}{D^4}\sum_{i=0}^{D^2}\sum_{j=0}^{D^2}U_iV_j\rho U_i^{\dagger}V_j^{\dagger},\\
        &= \sqrt{s (1-s)} \left( \dfrac{q_Aq_B}{D^2}\rho + q_A(1-q_B)\dfrac{I}{D} + q_B(1-q_A)\dfrac{I}{D} + (1-q_A)(1-q_B)\rho\right)\\
    {A}^{11} &= (1-s)\dfrac{q_Aq_B}{D^4} \sum_{i=0}^{D^2}\sum_{j=0}^{D^2}V_jU_i\rho U_i^{\dagger}V_j^{\dagger} = (1-s)\cN_B\cN_A(\rho).
\end{align}
\end{widetext}

\textbf{Nonlinear Channel Equalization. }
In wireless communication, signals sent from the antenna of a transmitter are transmitted to a receiver by following various paths while being reflected by structures such as buildings. Consequently, the transmitted signal is received with distortion due to the influence of noise added during transmission and the difference in transmission time depending on the path. Since this distortion depends on the frequency (channel), it is necessary to remove the distortion using an equalizer to demodulate the signal at the receiver. This process is called channel equalization.

In our numerical experiments, the distorted input $\{u_l\}$ is transformed from $\{s_l\}$ via the linear channel
\begin{align}
q_l&=0.08s_{l+2} - 0.12s_{l+1} + s_l + 0.18s_{l-1} - 0.1s_{l-2} \\
&+ 0.09s_{l-3} - 0.05s_{l-4} + 0.04s_{l-5} + 0.03s_{l-6} + 0.01s_{l-7}\nonumber,
\end{align}
and the nonlinear channel
\begin{align}
    u_l = q_l + 0.036q_l^2 - 0.011q_l^3 + \nu_l, 
\end{align}
where $\nu_l$ is an i.i.d. Gaussian noise~\cite{jaeger:2004:harness}.
We consider $\nu_l$ with zero mean adjusted in power to yield a signal-to-noise ratio as 24 dB.

\textbf{Training the Quantum Readout. }
In the classical readout, the training process is simply a linear regression of measurement results to target data, such as the tomography of the quantum state.
However, it is more complicated in the quantum readout since the target is the physical quantum state. An advantage of PRC is that we can keep the inner parameter fixed and train readout parameters and the interaction between the reservoir and the input.

In the quantum readout, we rely on the fact that any unitary matrix that describes the mixing between optical modes can be implemented with linear optics devices such as phase shifters and beam spliters~\cite{braunstein:2005:toolbox}.
Therefore, we can implement the combination of the transmitted fields to generate $M$ quantum output modes 
$
    \hat{C}_m = \sum_j o_{mj}\hc_j
$ with complex coefficients $o_{mj}$.
The output modes must satisfy the commutation relations $[\hat{C}_m, \hat{C}_n^{\dagger}] = \delta_{mn}$, which impose the condition $\sum_j o_{mj}o^{*}_{nj}=\delta_{mn}$.

Let us consider $\boldsymbol{\theta}$ as the vector of transformed real parameters for the interaction coefficients $\{W^{\textup{in}}_{jk}\}$ and readout coefficients $\{o_{mj}\}$.
Given $L$ training data with hybrid inputs $\{(u_l, \beta_l)\}$ and target quantum states $\hsig_l$, the cost function 
$\cL_{\boldsymbol{\theta}}(\{(u_l, \beta_l), \hsig_l\}_{l=1}^{L})$ evaluates the difference between the quantum states $\{\sigma_l\}_{l=1}^{L}$ described via $\{\hat{C}_m\}$ and the target quantum states $\{\hsig_l\}_{l=1}^{L}$.
In our numerical simulations, $\cL_{\boldsymbol{\theta}}$ is defined via the fidelity error [Eq.~\eqref{eqn:fidelity:error}] or the error taken in the Wigner representation [Eq.~\eqref{eqn:wigner:error}].
Here, $\cL_{\boldsymbol{\theta}}$ becomes a nonlinear function of parameters $\boldsymbol{\theta}$. We find the optimal $\boldsymbol{\theta}$ such that $\cL_{\boldsymbol{\theta}}$ is minimized. 
Several methods can be used for this nonlinear optimization problem, and we use the Nelder\text{--}Mead simplex algorithm~\cite{lagarias:1998:nd}, which is fast and effective for problems with a large number of parameters.
The algorithm starts from an initial guess for the parameters and generates a simplex in the multidimensional parameter space.
In each iteration, the cost function is evaluated at each point in the simplex. 
Under a selecting and replacing procedure, the points in the simplex with the worst value of the cost function are reconstructed for each step until a convergence condition is satisfied.
In our simulation, we use the Julia framework~\cite{bezanson:2017:julia} for simulating the quantum master equation and the built-in function with default parameters in the Optim package for the Nelder\text{--}Mead algorithm.
We refer to Supplementary Information for more detailed results obtained using the Nelder\text{--}Mead steps in the optimization process.

\begin{acknowledgments}
The authors acknowledge Shumpei Kobayashi for fruitful discussions.
This work is supported by MEXT Quantum Leap Flagship Program (MEXT Q-LEAP) Grant Nos. JPMXS0118067394 and JPMXS0120319794.
\end{acknowledgments}

\section*{Author contributions}
All authors conceived the research and contributed significantly to interpreting the results. 
Q.H.T conceived the model, performed the main analysis for the experimental data and prepared the manuscript.
S.G. developed the concept and designed the initial simulation for quantum reservoir.
K.N. supervised the research and contributed to the ideation and design of the research. 
All authors contributed to writing the manuscript.

\section*{Competing interests}
The authors declare no competing interests.

\section*{Additional information}
\textbf{Supplementary information} The online version contains supplementary material.

\textbf{Correspondence} and requests for materials should be addressed to Q. H. Tran.

\end{document}


\title{Supplementary Material for\\
		``Quantum\text{--}Classical Hybrid Information Processing \\via a Single Quantum System''}
	
	\author{Quoc Hoan Tran}
	\email{k09tranhoan@gmail.com}
	\affiliation{Next Generation Artificial Intelligence Research Center (AI Center), Graduate School of Information Science and Technology, The University of Tokyo, Japan}
	
	\author{Sanjib Ghosh}
	\email{sanjibghosh@baqis.ac.cn}
	\affiliation{
		Beijing Academy of Quantum Information Sciences, Beijing, China
	}
	
	\author{Kohei Nakajima}
	\email{k-nakajima@isi.imi.i.u-tokyo.ac.jp}
	\affiliation{Next Generation Artificial Intelligence Research Center (AI Center), Graduate School of Information Science and Technology, The University of Tokyo, Japan}

	\date{\today}
	
	\begin{abstract}
		This supplementary material describes in detail the calculations, the experiments introduced in the main text, and the additional figures.
		The equation, figure, and table numbers in this section are
		prefixed with S (e.g., Eq.\textbf{~}(S1) or Fig.~S1, Table~S1),
		whereas numbers without the prefix (e.g., Eq.~(1) or Fig.~1, Table~1) refer to items in the main text.
	\end{abstract}
	
	\maketitle
	\tableofcontents
	\newpage
	
	\section{Learning Quantum Tomography via Quantum Reservoir Computing}
	In this section, we explain some quantum extensions of reservoir computing (RC) using a scheme called quantum reservoir computing (QRC) for the quantum tomography task.
	
	\subsection{Quantum Reservoir Computing with Network of Quantum Dots}

	In our study, we model the QRC approach via the framework of repeated quantum interactions.
	Here, the input sequence is fed via sequential interactions between a quantum reservoir (QR) network $\cS$ with an auxiliary system $\cE$.
	We consider a two-dimensional lattice of $N$ quantum dots for the QR network, represented by the Hamiltonian
	\begin{align}\label{eqn:hamiltonian}
		\hat{H} = &\sum_i E_i\hcd_i\hc_i + \sum_{\langle i,j \rangle}h_{ij}\left( \hcd_i\hc_j +  \hcd_j\hc_i\right) + \sum_iQ_i\hcd_i\hcd_i\hc_i\hc_i + P(t)\sum_i\left( \hcd_i + \hc_i \right),
	\end{align}
	where $\hc_i$, $E_i$, $h_{ij}$, $Q_i$, and $P(t)$ are the field operators, onsite energies, hopping amplitudes between the nearest neighbor sites, nonlinearity strengths, and uniform time-dependent coherent field strengths to excite the QR, respectively. 
	
	The dynamical evolution can be described by the master equation (we omit the Plank constant $\hbar$ for ease notation).
	\begin{align}
		\dot{\rho} = -i[\hat{H}, \rho] + \gamma\sum_j\cL(\hc_j)\rho + \Omega(t - t_{\text{init}})\hat{A}\rho,
	\end{align}
	where $\rho$ is the combined density matrix of the QR as well as the input modes.
	Here, $\Omega(t)=1$ for $t\geq 0$ and $0$ otherwise,
	and $\hat{A}\rho=\sum_k \dfrac{\gamma_k}{\gamma}\cL(\ha_k)\rho + \sum_{k,j}W_{jk}^{\text{in}} \left( \left[\ha_k\rho, \hcd_j \right] + \left[\hc_j, \rho\had_k \right] \right)$ represents the decay in the input modes with rates $\gamma_k/\gamma$ (the first term)
	due to the cascaded coupling between the input modes $\ha_k$ and the QR (the remaining terms). 
	The Lindblad superoperator $\cL(\hat{x})$ is defined for any arbitrary operator $\hat{x}$ by $\cL(\hx)\rho=\hx\rho\hxd - \dfrac{1}{2}\left( \hxd\hx\rho + \rho\hxd\hx\right)$.
	In our numerical simulations, we consider $E_i/\gamma=0$, $\gamma_k/\gamma = \sum_j (W_{jk}^{\text{in}})^2$, with $W_{jk}^{\text{in}}$ being the input weights randomly chosen from the interval $[0.0, \gamma]$.
	For the tomography tasks, we assume that there is no interaction between reservoir sites, i.e., $Q_i=0$ (Figs. 2,3,4 in the main text).
	We allow for the interactions ($Q_i/\gamma = 1.0$) in the experiments preparing the quantum depolarizing channel (Fig.~5 in the main text).
	
	In the information processing pipeline, the QR is first excited with a uniform $P(t)=P$ for $0\leq t< t_{\text{init}}$ and no incident quantum inputs. We choose $t_{\textup{init}}$ such that the QR at time $t_{\text{init}}$ reaches a steady state.
	Then, the quantum input is incident to the reservoir via the input modes $\ha_k$, and the classical input $u(t)=u$ is activated at the same time.
	The classical input $u(t)$ is encoded via the classical optical excitation as 
	$P(t) = P + Wu(t)$, where $W$ is the input scaling.
	At time $t_1 = t_{\textup{init}} + \tau$ for the time interval $\tau$, an appropriate and practical readout from the QR is performed for nontrivial transformations of input data, such as a linear combination of quantum modes or the linear combination of measurement on the accessible observables.
	If the task is a non-temporal processing task, we repeat the above procedure for every hybrid data instance $(u, \beta)$.
	For a temporal processing task, we consider a sequence of hybrid inputs $(u_1, \beta_1), (u_2, \beta_2), \ldots$, where $\{u_l\}$ is the classical and $\{\beta_l\}$ is the quantum sequence. 
	At time $t_l = t_{\textup{init}} + (l-1)\tau$ for $l=1,2,\ldots$, the classical input is switched to $u(t)=u_l$, and the quantum state $\beta_l$ replaces the partial state in the input modes.

	In a design similar to that of classical RC, we first consider a readout scheme called \textit{classical readout} associated with a measurement process. The partial information regarding the state of the QR network after the $n$th interaction is obtained via measuring the expectation values of the occupation numbers in reservoir sites.
	Here, the $j$th element $x_{lj}$ of the reservoir states $\bx_l$ can be calculated as $x_{lj}= n_j=\langle \hcd_j\hc_j\rangle$.
	One can increase the dimension of $\bx_l$ by using the temporal multiplexing technique.
	Between two inputs, we perform measurements at every time interval $\tau/V$, where
	$V$ is the measurement multiplexity. 
	After obtaining the reservoir states, the training procedure in QRC is similar to that in conventional RC.
	In the classical readout, multitasking is possible because the training cost is minimal for independent training with different readout weights for different tasks.
	
\begin{figure}
		\includegraphics[width=12cm]{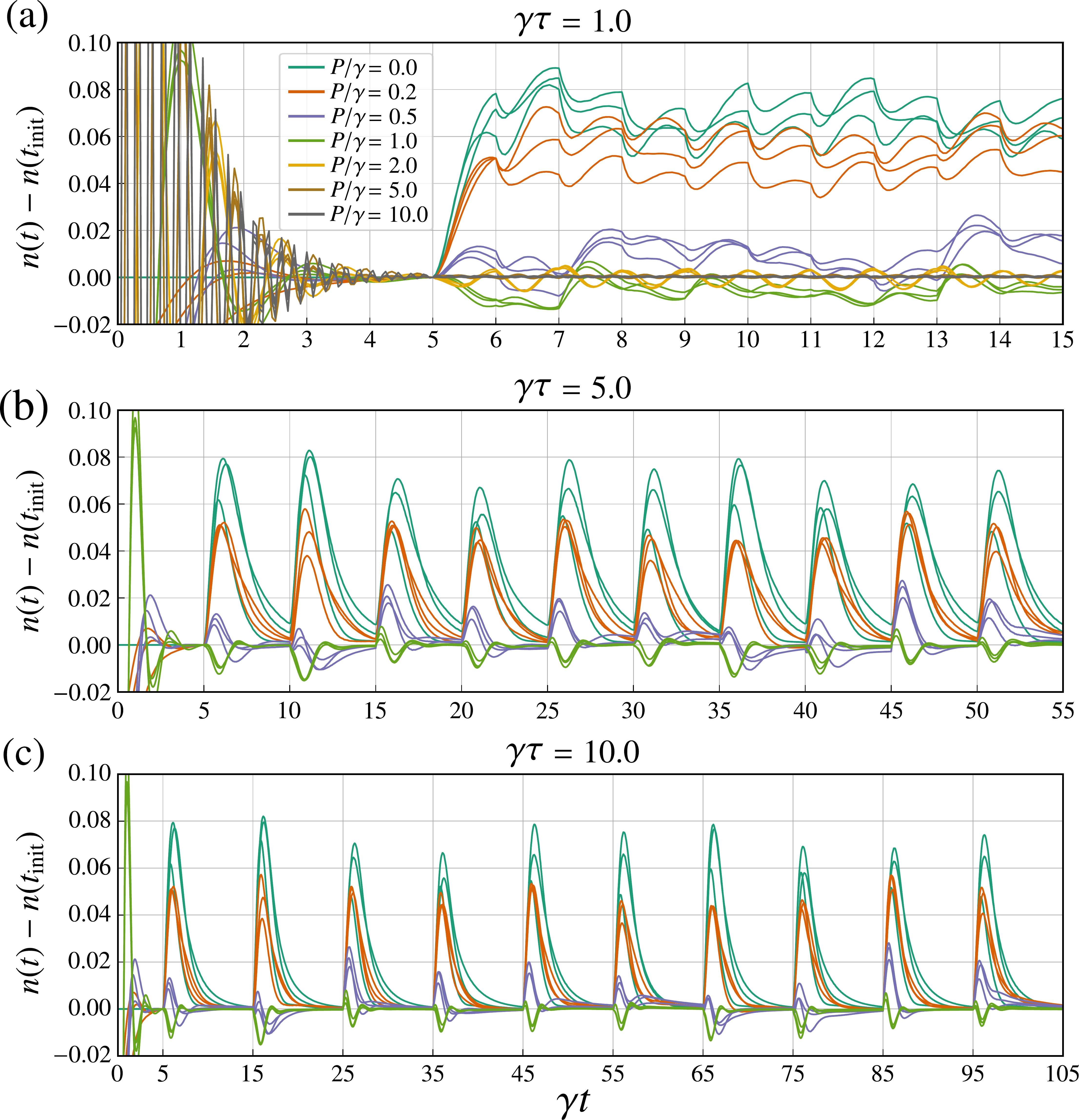}
		\protect\caption{The typical dynamics of the occupation numbers $n_j(t)$ in the QR, represented by the amount of the input photons $n_j(t) - n_j(t_{\textup{init}})$ entering the QR. The dynamics starts from $t=0$ to $t_{\textup{init}}=5.0/\gamma$, where one-qubit quantum inputs are incident at $t_l = t_{\textup{init}} + (l-1)\tau$. 
			\label{fig:tau}}
	\end{figure}

	The learning performance of the classical readout scheme depends on the dynamics of the occupation numbers $n_j(t)$. In Fig.~\ref{fig:tau}, we show the dynamics of the occupation numbers $n_1(t)$, $n_2(t)$, and $n_3(t)$ compared with the corresponding numbers at time $t_{\textup{init}}$ for $N=3$ reservoir sites with a constant classical input ($W/\gamma=0$) and randomized quantum inputs of one-qubit quantum states. Here, we consider $t_{\textup{init}}=5.0/\gamma$ with different input intervals $\tau$ and different incoherent excitation $P$. 
	The dynamics from $t=0$ to $t_1 = 5.0/\gamma$ is solely driven by $P$ (actually, the dynamics starts with an empty reservoir $n_1(t)=n_2(t)=n_3(t)=0$) and the system reaches to an initial state before $t_{\textup{init}}$.
	The first quantum input is incident at $t_1 = t_{\textup{init}}$, where we can see that the occupation numbers deviate from the steady states.
	If we consider a large $\tau$ [Figs.~\ref{fig:tau}(b)(c)], the occupation numbers come back to the steady state before the next input. If we perform the readout measurements at this timing, we cannot obtain sufficient information from the input. Therefore, we choose $\tau$ such that $n_j(t)$ are largely away from the steady values at the timing before the next input is incident on the system [Fig.~\ref{fig:tau}(a)]. In our experiments, we consider $\gamma\tau = 1.0$.
	
	\subsection{Channel Equalizer and the Tomography of the Quantum Switch}
	As a demonstration in the main text, we consider the quantum switch, which is a superposition of two alternative orders of quantum channels~\cite{chiribella:2012:qswitch}.
	A quantum switch includes two quantum channels $\cN_A$ and $\cN_B$ to create a new channel $\cS(\cN_A, \cN_B)$, which uses the channels $\cN_A$ and $\cN_B$ in an order that is entangled with an independent switch quantum state $\rho_s$. 
	The channel $\cS(\cN_A, \cN_B)$ returns the state $(\cN_A\circ\cN_B(\rho))\otimes \ket{0}\bra{0}$ if $\rho_s = \ket{0}\bra{0}$ and $(\cN_B\circ\cN_A(\rho))\otimes \ket{1}\bra{1}$ if $\rho_s = \ket{1}\bra{1}$.
	When the switch state is in a superposition of $\ket{0}$ and $\ket{1}$, the channel returns a correlated state as the result of $\cN_A$ and $\cN_B$ acting on $\rho$ in a quantum superposition of two alternative orders.

	We consider a random sequence of one-qubit input states $\{\beta_l\}$ 
	and the classical control $\{s_l\}$ as an i.i.d. discrete sequence of symbols, which are selected from $\{-3, -1, 1, 3\}$ with equal probability. The switch state at each $l$ is $\rho_{(3+s_l)/6}$, and the distorted input $\{u_l\}$ is transformed from $\{s_l\}$ via two channels: $q_l=0.08s_{l+2} - 0.12s_{l+1} + s_l + 0.18s_{l-1} - 0.1s_{l-2} + 0.09s_{l-3} - 0.05s_{l-4} + 0.04s_{l-5} + 0.03s_{l-6} + 0.01s_{l-7}$ (linear channel) and $u_l = q_l + 0.036q_l^2 - 0.011q_l^3 + \nu_l$ (nonlinear channel) where $\nu_l$ is an i.i.d. Gaussian noise.
	Given a delay $d$ and current inputs $\beta_l$, $u_l$, we aim to perform a tomography of the temporal quantum map $\cF(\{s_l, \beta_l\}_{l}) = \cS(\cN_A, \cN_B)(\beta_{l-d}\otimes \rho_{s_{l-d}})$ and reconstruct $s_{l-d}$, where $\cN_A, \cN_B$ are two depolarizing quantum channels.
	This reconstruction of $\{s_l\}$ is considered the nonlinear channel equalization task, which has been noted in wireless communication channels and demonstrated using conventional RC method~\cite{jaeger:2004:harness}.

	\begin{figure}
		\includegraphics[width=15.5cm]{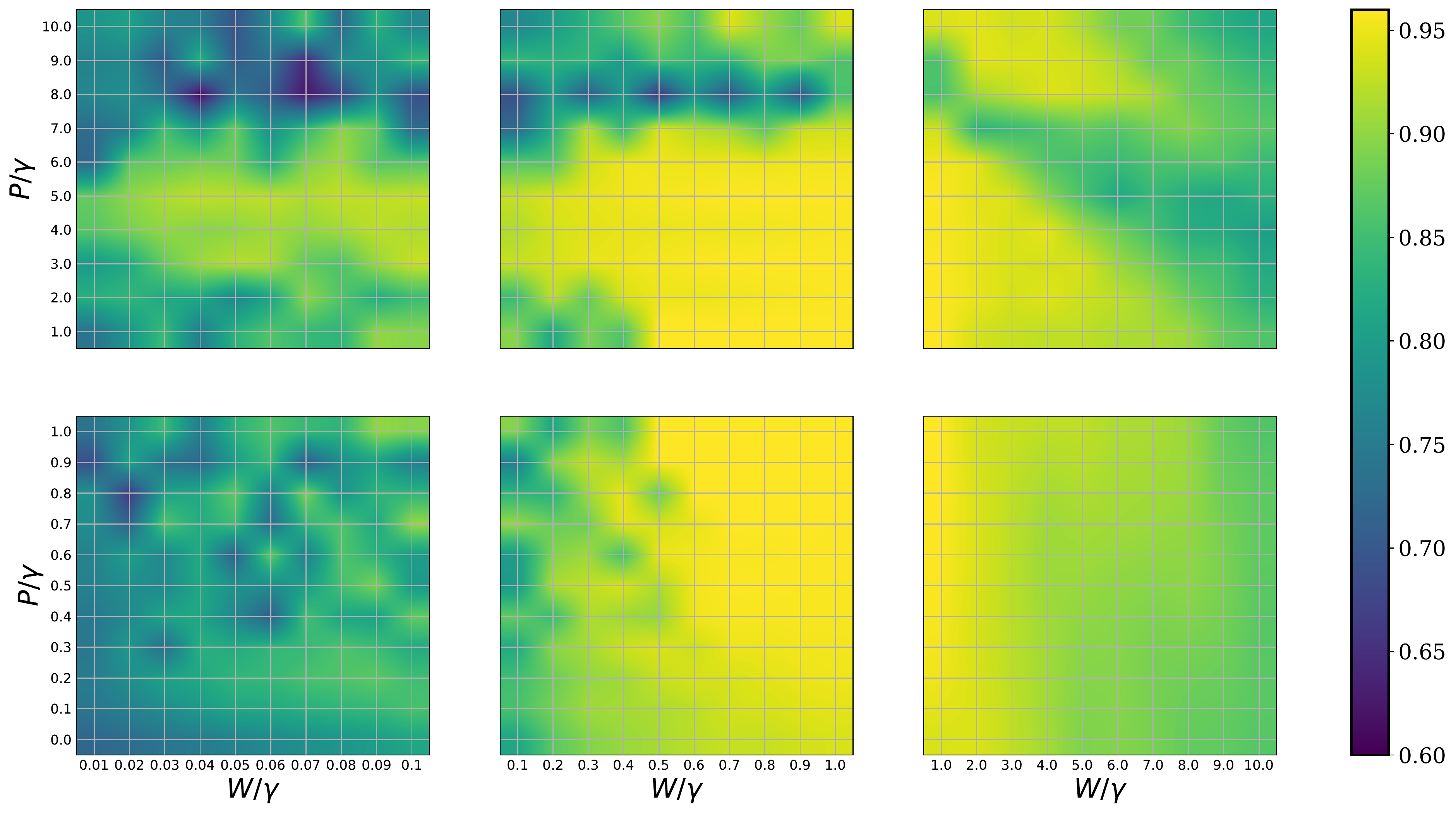}
		\protect\caption{The average root mean square of fidelities (RMSF) (the higher value is the better performance) over 10 trials of random input data, QR and quantum switch configurations at the combination of $P$ and $W$.  We use $N=3$ reservoir sites and a measurement multiplexity $V=8$.
			\label{fig:sw_eqn_quantum}}
	\end{figure}
	
	\begin{figure}
		\includegraphics[width=15.5cm]{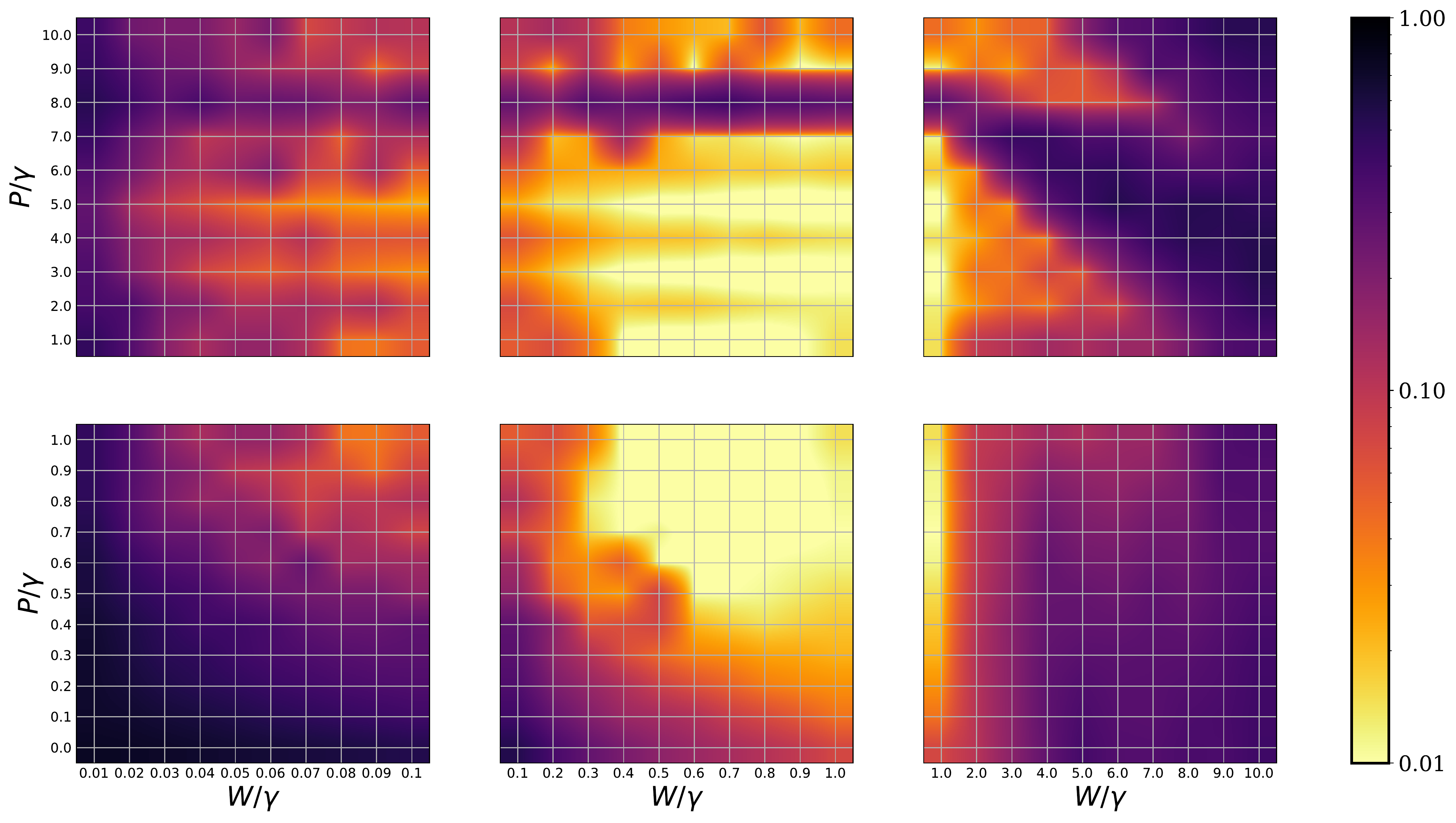}
		\protect\caption{The average symbol error rate (SER) (the lower value is the better performance) over 10 trials of random input data, QR's configuration at the combination of $P$ and $W$. We use $N=3$ reservoir sites and the measurement multiplexity $V=8$.
			\label{fig:sw_eqn_class}}
	\end{figure}
	
	Figures~\ref{fig:sw_eqn_quantum} and~\ref{fig:sw_eqn_class} present the average performance over 10 trials for each combination of $P/\gamma$ and input scaling $W/\gamma$ with $N=3$ reservoir sites, where in each trial, we consider a different random input sequence, random QR configurations, and random depolarizing coefficients.
	The training is performed at $l=1,\ldots,L$, and the performance is evaluated over $l=L+1,\ldots,L+T$ via the root mean square of fidelities (RMSF) in the tomography task $\textup{RMSF}=\sqrt{(1/T)\sum_{l=L+1}^{l=L+T}F^2(\sigma_l, \hat{\sigma}_l)}$,
	and the symbol error rate (SER) $\text{SER}=\textup{card}(\{l \mid s_{l-d} \neq \hat{s}_l\})/T$. Here, the fidelity $F(\rho,\sigma )=\tr[\sqrt{\sqrt{\sigma}\rho\sqrt{\sigma}}]$ and $\sigma_l$ and $s_{l-d}$ are the targets for the tomography and channel equalization tasks, respectively.
	We use 800 time steps for training and 200 time steps for the evaluation.
	We also use the  temporal multiplexing technique with the measurement multiplexity V = 8 for numerical experiments.

\begin{figure}
		\includegraphics[width=17cm]{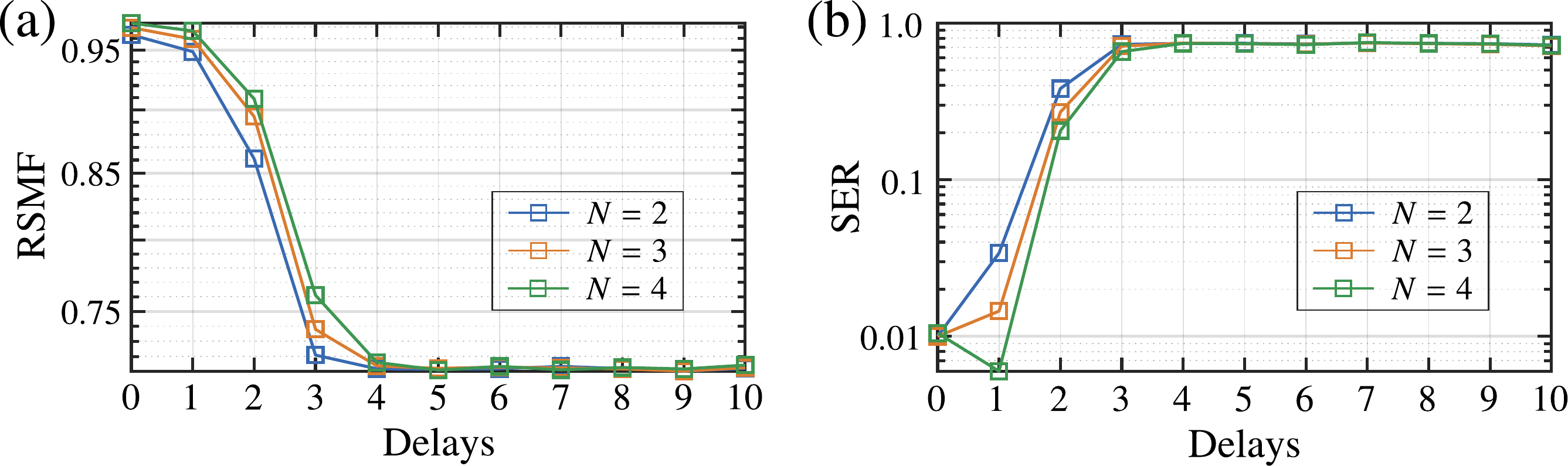}
		\protect\caption{The average root mean square of fidelities (RMSF) of the tomography task (a) and the average symbol error rate (SER) of the channel equalization task (b) over 10 trials when we increase the value of the delay $d$.  
			Here, we consider the QR with $N=2, 3, 4$ sites in the reservoir.
			\label{fig:sw_eqn_delays}}
	\end{figure}

	Figure~\ref{fig:sw_eqn_delays} displays the RMSF of the tomography task and the SER of the channel equalization task when we increase the value of the delay $d$ with different number of sites in the QR.  
	The performances drop significantly when $d\geq 3$, implying the short-term memory property in the QR.
	
	\subsection{Non-temporal Continuous Variable Tomography}
	
	In this demonstration, we consider the quantum tomography of continuous variable states in a non-temporal setting. 
	We evaluate the tomography for three settings: $\cF_{\textup{cv-amp}}, \cF_{\textup{cv-phase}}$, and $\cF_{\textup{cv-sw}}$, of the quantum map $\cF(s, \beta)$, given an one-mode quantum input $\beta$ and a classical input $s$.
	
	For $\cF_{\textup{cv-amp}}$ and $\cF_{\textup{cv-phase}}$, we consider a random sequence in $[0, 1]$ of $\{s_l\}_{l=1:200}$  and a random sequence of one-mode thermal states $\{\beta_l\}_{l=1:200}$ and take the index of $l=1,\ldots,100$ for the training and $l=101,\ldots,200$ for the evaluation. Here, we consider one-mode thermal states $\beta_l$ as Gaussian continuous-variable states with the density matrices
	\begin{align}\label{eqn:thermal}
		\beta_l = \sigma_l \quad \text{ for } \quad \sigma_l = \dfrac{1}{1 + \overline{v}_l}\sum_{n=0}^{\infty}\left( \dfrac{\overline{v}_l}{1 + \overline{v}_l}\right)^n \ket{n}\bra{n},
	\end{align}
	where $\ket{n}$ represents the state corresponding to $n$ photon numbers, and $\overline{v}_l$ is the expectation value of the photon number in the state. We consider $\overline{v}_l = (r_l \cos(\phi_l))^2$, where $r_l$ and $\phi_l$ are taken randomly in $[0.0, 0.3]$ and  $[0.0, \pi]$, respectively.
	The quantum maps $\cF_{\textup{cv-amp}}$ and $\cF_{\textup{cv-phase}}$ are defined as
	\begin{align}
		\cF_{\textup{cv-amp}}(s, \beta) &= \hat{\cS}(\xi_\textup{amp}(s))\beta \hat{\cS}(\xi_\textup{amp}(s))^{\dagger},\\
		\cF_{\textup{cv-phase}}(s, \beta) &= \hat{\cS}(\xi_\textup{phase}(s))\beta \hat{\cS}(\xi_\textup{phase}(s))^{\dagger},
	\end{align}
	where $\cS(\xi)$ is the one-mode squeezing operator, defined as $\cS(\xi) = \exp(\xi \had\had - \xi^*\ha\ha)$.
	Here, we consider $\xi$ as functions of classical data $s$ defined as
	\begin{align}
		\xi_\textup{amp}(s) = s\exp(i\pi/4),  \qquad \xi_\textup{phase}(s) = 0.3\exp(i2\pi s).
	\end{align}
	
	For the quantum map $\cF_{\textup{cv-sw}}$, we consider the same$\{s_l\}$ but random one-mode squeezed-thermal states for $\{\beta_l\}$ as
	\begin{align}
		\beta_l = \hat{\cS}(\xi_l)\sigma_l\hat{\cS}(\xi_l)^{\dagger},
	\end{align}
	where $\sigma_l$ is defined as in Eq.~\eqref{eqn:thermal} and $\xi_l = r_l\sin(\phi_l)$.
	The quantum map $\cF_{\textup{cv-sw}}$ is defined as the quantum switch with the input $\beta_l$ and the switch state \begin{align}
		\psi_{s_l} = \sqrt{s_l}\ket{\alpha} + \sqrt{1-s_l}\ket{-\alpha},
	\end{align}
	where we consider the following coherent states with $\alpha = 2.5$
	\begin{align}
		\ket{\pm\alpha} = \exp\left(-\dfrac{|\alpha|^2}{2} \right)\sum_{n=0}^{\infty}\dfrac{(\pm\alpha)^n}{\sqrt{n!}}\ket{n}.
	\end{align}
	
	\begin{figure}
		\includegraphics[width=15cm]{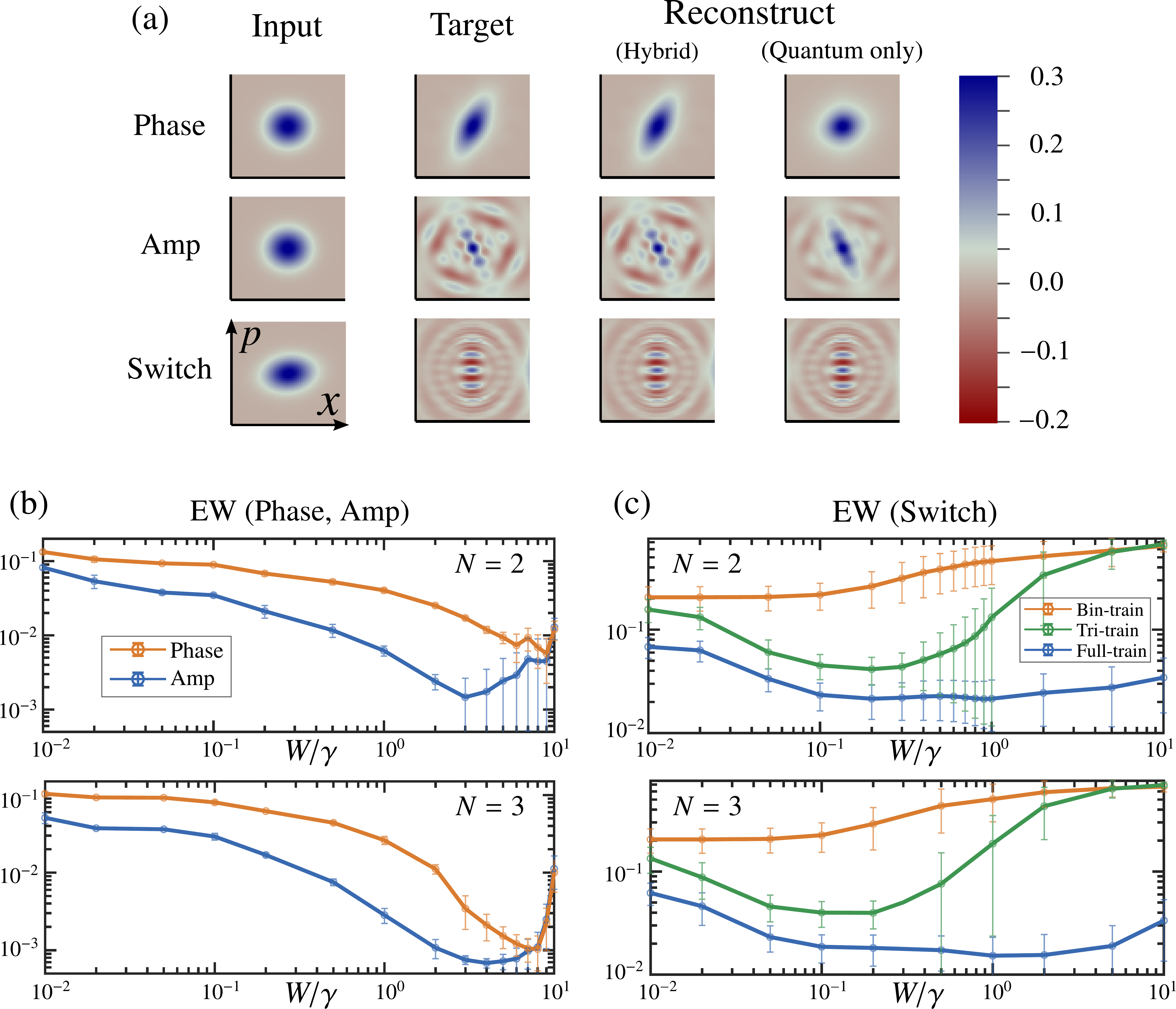}
		\protect\caption{(a) Examples of input, target, and reconstruct Wigner functions in the non-temporal continuous variable tomography for $\cF_{\textup{cv-phase}}$ (upper panel), $\cF_{\textup{cv-amp}}$ (middle panel), and $\cF_{\textup{cv-sw}}$ (lower panel) by $N=4$-site QR with a measurement multiplexity $V=10$ and a constant coherent field strength $P/\gamma=1.0$. 
			The columns labeled ``(Hybrid)" and ``(Quantum only)" describe the results with the input scaling $W/\gamma=1.0$ (both classical and quantum inputs) and $W/\gamma=0.0$ (no classical input), respectively. 
			(b) Variation in the tomography error EW for $\cF_{\textup{cv-phase}}$ (orange line) and $\cF_{\textup{cv-amp}}$ (blue line) with the input scaling $W$ and $N=2, 3$-site QR.
			(c) Variation in the tomography error EW for $\cF_{\textup{cv-sw}}$ and $N=2, 3$-site QR.
			In the results labeled as ``Full-train",  we consider a random sequence in $[0, 1]$ of $\{s_l\}_{l=1:200}$  and a random sequence of one-mode thermal states $\{\beta_l\}_{l=1:200}$ with the index of $l=1,\ldots,100$ for the training and $l=101,\ldots,200$ for the evaluation. 
			In the results labeled ``Bin-train" and ``Tri-train", only binary or tri-values classical data in the training phase are considered, i.e., $s_l\in\{0.0, 1.0\}$ (for ``Bin-train") and $s_l \in \{0.0, 0.5, 1.0\}$ (for ``Tri-train") for $l=1,\ldots,100$.
			\label{fig:qlm:demo}}
	\end{figure}
	
	Figure~\ref{fig:qlm:demo}(a) shows several examples of input, target, and reconstruct Wigner functions for $\cF_{\textup{cv-phase}}$ (upper panel), $\cF_{\textup{cv-amp}}$ (middle panel), and $\cF_{\textup{cv-sw}}$ (lower panel). Here, we use $N=4$-site QR with a measurement multiplexity $V=10$ and a constant coherent field strength $P/\gamma=1.0$. 
	The columns labeled ``(Hybrid)" and ``(Quantum only)" describe the results when we consider the input scaling $W/\gamma=1.0$ (both classical and quantum inputs) and $W/\gamma=0.0$ (no classical input), respectively. 
	We observe that if both of quantum and classical inputs are included in the QR, the reconstructed states are very similar to the target states of the quantum maps with hybrid quantum-classical input.
	
	To further evaluate the performance systematically, we calculate the error based on the Wigner representation
	\begin{align}
		\textup{EW} = \sqrt{ \dfrac{1}{T}\sum_{l=L+1}^{L+T}\dfrac{\sum_{i,j}[\cW(\cF_l; x_i, p_j) - \hat{\cW}(\cF_l; x_i, p_j)]^2}{\sum_{i,j}[\cW(\cF_l; x_i, p_j) + \hat{\cW}(\cF_l; x_i, p_j)]^2}},
	\end{align}
	where $\cW(\cF_l; x_i, p_j)$ and $\hat{\cW}(\cF_l; x_i, p_j)$ are the target and reconstructed Wigner functions of the state $\cF_l = \cF(s_l, \beta_l)$, respectively. The error metric is evaluated in the evaluation phase with $L=100$ and $T=100$.
	
	Figure~\ref{fig:qlm:demo}(b) depicts the errors to reconstruct the quantum maps $\cF_{\textup{cv-phase}}$ (orange lines) and $\cF_{\textup{cv-amp}}$ (blue lines) at different input scaling $W$ with $N=2$ (upper plot) and $N=3$ (lower plot).
	The errors are calculated over 10 trials of data and QR configurations, with the solid lines depicting the average values associated with error bars.
	We observe an optimal range of input scaling $W$ for optimal performance in each task.
	We note that setting a too small value of $W$ limits the effect of the classical input into the QR.
	In contrast, setting a too large value of $W$ will impose the localization in the quantum dots and may lead to non-ergodic behavior in the QR.
	In this case, when the input state $\beta_l$ is incident on the QR with weak coupling ($|W_{jk}^{\textup{in}}| \ll |P + Ws_l|$), sufficient information regarding $\beta_l$ cannot be extracted from the QR.
	
	Figure~\ref{fig:qlm:demo}(c) depicts the errors $\textup{EW}$ (blue lines labeled ``Full-train") to reconstruct the quantum map $\cF_{\textup{cv-sw}}$. We can observe that the effect of input scaling $W$ is not significant as in other tasks. 
	We further present an intriguing setting by limiting the variety of the classical input in the training phase while keeping the same data in the evaluating phase.
	In the results labeled ``Bin-train", we only consider binary classical data in the training phase, i.e., $s_l\in\{0.0, 1.0\}$ for $l=1,\ldots,100$.
	The performance is worse since there is no superposition of the switch state in the training phase to help the learning of the quantum switch.
	However, for tri-values of $s_l \in \{0.0, 0.5, 1.0\}$ (labeled ``Tri-train") in the training phase, only one pattern of the superposition switch state is considered in the training; we can obtain a relatively low error with a suitable range of input scaling $W$. For example, the performance at $W=0.2$ is comparable with the performance of ``Full-train" at $W=0.02$.
	These results demonstrate that tomography for the quantum switch can be performed with limited patterns of training data.
	
	
	\section{Continuous Variable Tomography and Closed-Loop Control}
	
\begin{figure}
		\includegraphics[width=12cm]{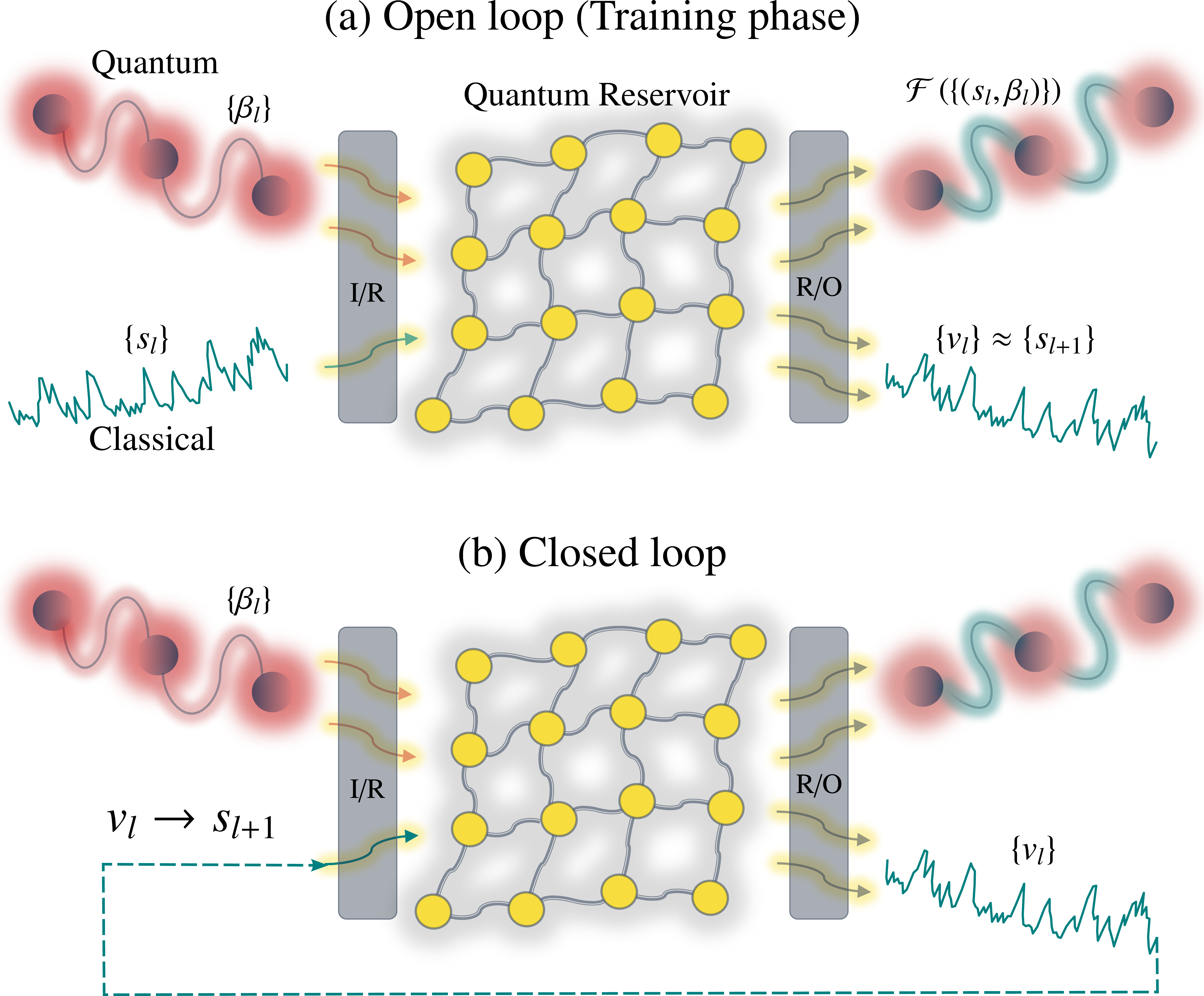}
		\protect\caption{Closed-loop control of the QR. The QR can autonomously generate classical data $s_l$ in a closed-loop manner while performing the tomography task.
			(a) In the training phase, $s_l$ is learned in an open loop to output the classical value $v_l$ given the input $s_l$ such that $v_l$ and $s_{l+1}$ are as close as possible.
			(b) After training, the prediction is used as the classical input for the next step, i.e., replacing $s_{l+1}$ by $v_l$, thereby forming a closed-loop control without any external interventions.
			\label{fig:close:overview}}
	\end{figure}

	In this section, we formulate an intriguing problem in predicting the future evolution of the quantum tomography of hybrid inputs. Specifically, we consider a situation where after the training phase, we cannot access the information from the classical control $s_l$.
	This problem can be addressed owing to the advantage of the multitasking in reservoir computing.
	Here, our QR can autonomously generate $s_l$ in a closed-loop manner while performing the tomography task using hybrid inputs.
	In the training phase, $s_l$ is learned in open loop, where we output the classical value $v_l$ given the input $s_l$ such that $v_l$ and $s_{l+1}$ are as close as possible [Fig.~\ref{fig:close:overview}(a)].
	After training, the prediction is used as the classical input for the next step, i.e., replacing $s_{l+1}$ by $v_l$, thereby forming a closed-loop control without any external interventions [Fig.~\ref{fig:close:overview}(b)].
	
	In this demonstration, we consider the quantum tomography of continuous variable states. The target is to reconstruct the representation of the output $\cF_l= \cF\{(s_l,\beta_l)\}$ in the Wigner function form $\cW(\cF_l; x_i, p_j)$ defined on a grid of continuous variables $x_i$ and $p_j$.
	We use 300 randomly generated one-mode thermal states $\beta_l$ and the periodic signals $s_l = 0.5 + 0.5\sin(\dfrac{l\pi f}{510})$ as the hybrid input in the training phase.
	The target squeezing thermal state $\cF_l$ is created by applying the one-mode squeezing operator $\hat{S}(\xi_l)=\exp\left( \xi_l \had\had - \xi_l^{*}\ha\ha \right)$ to $\beta_l$.
	We consider two types of output squeezing thermal states, where the classical control
	$s_l$ is encoded in the amplitude ($\xi_l=s_le^{i\pi/4}$) or phase ($\xi_l=0.3e^{i2\pi s_l}$) of $\xi_l$.
	Here, we consider the cutoff Fock space dimension (the effective dimension) of these continuous variables states is $D_{\textup{eff}}=9$.
	
\begin{figure}
		\includegraphics[width=16cm]{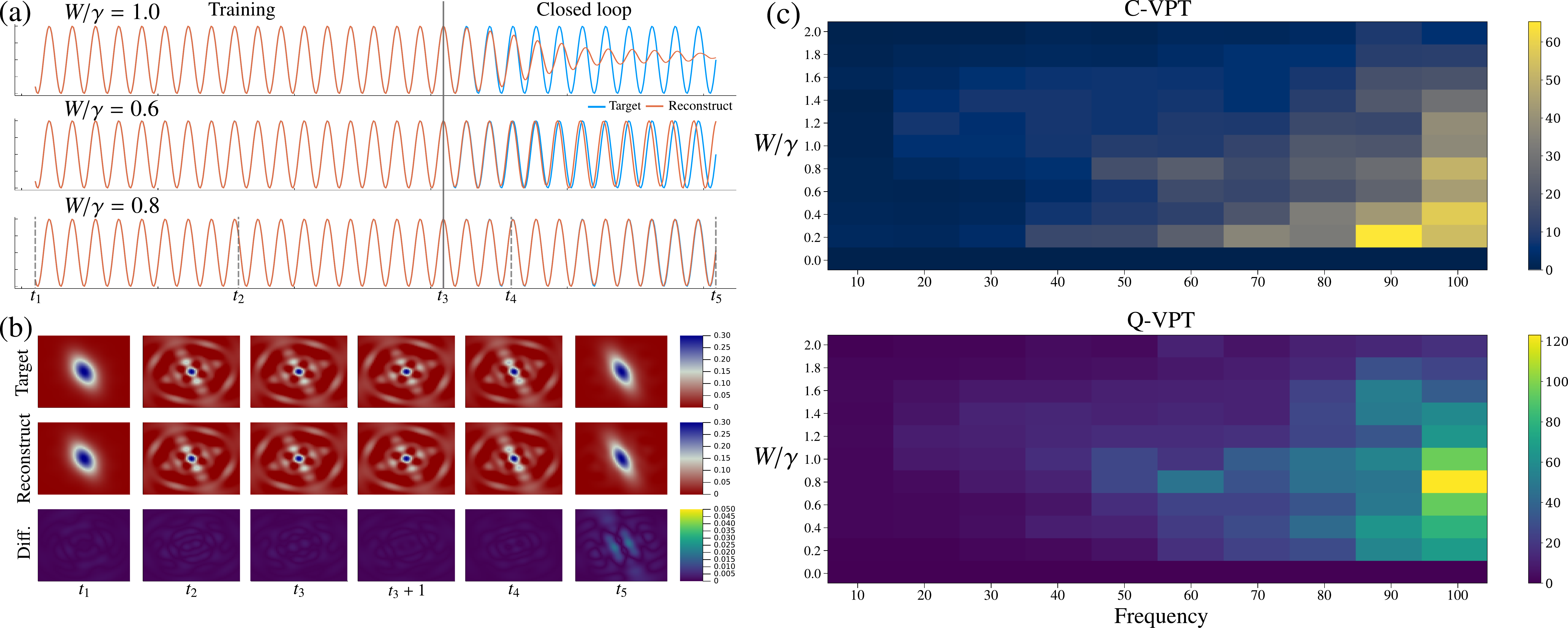}
		\protect\caption{(a) Examples of the classical control signals in the training and closed-loop phase for $f=60$ with different values of the input scaling $W/\gamma$ in the QR of $N=3$ sites with measurement multiplexity $V=10$.
			(b) Continuous variable tomography at typical time steps in the training phase ($t_1, t_2, t_3$) and the closed-loop phase ($t_3+1, t_4, t_5$)  with $W/\gamma=0.8$ in the amplitude encoding of the classical control in the target states. The last panel displays the absolute difference between the target and reconstructed Wigner functions.
			(c) The valid prediction time in classical (C-VPT) and quantum tomography (Q-VPT) tasks depending on the input scaling $W/\gamma$ and the frequency of the sinusoidal classical inputs.
			\label{fig:close:amp}}
	\end{figure}

\begin{figure}
		\includegraphics[width=16cm]{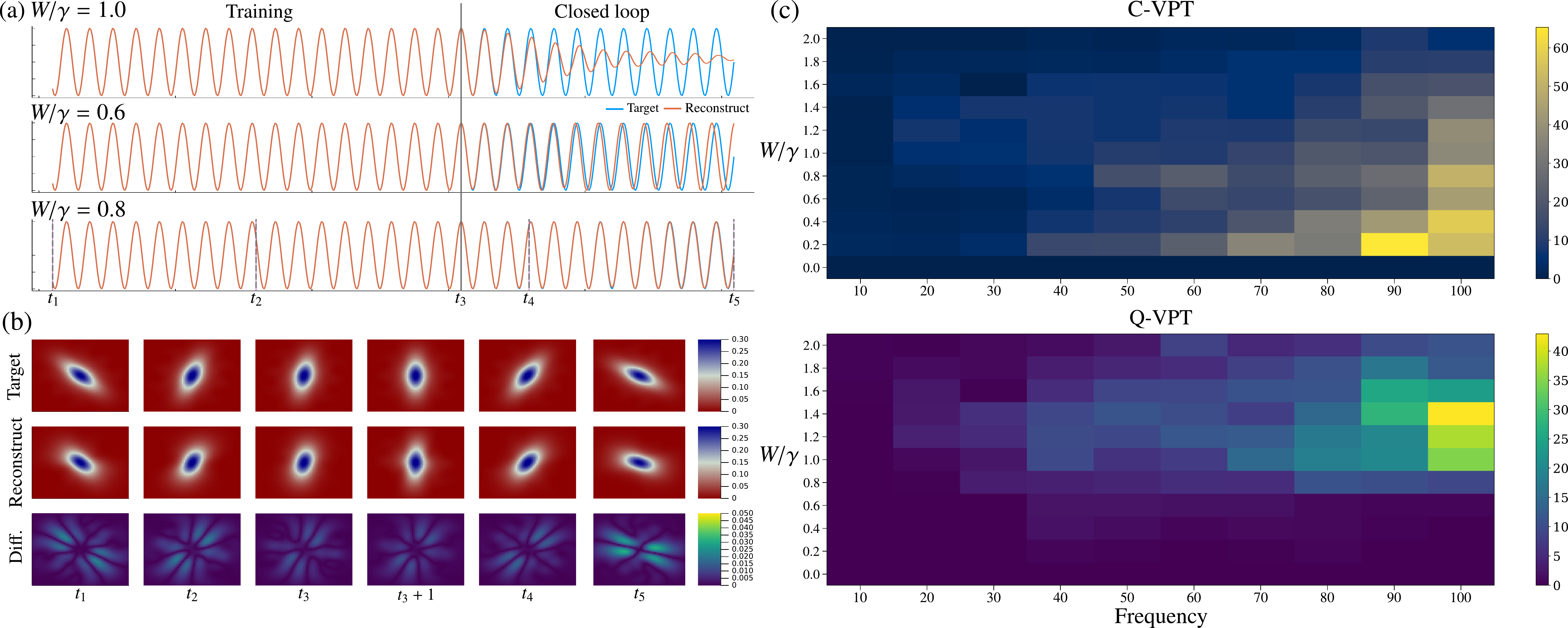}
		\protect\caption{(a) Examples of the classical control signals in the training and closed-loop phase for $f=60$ with different values of the input scaling $W/\gamma$ in the QR of $N=3$ sites with measurement multiplexity $V=10$.
			(b) Continuous variable tomography at typical time steps in the training phase ($t_1, t_2, t_3$) and the closed-loop phase ($t_3+1, t_4, t_5$)  with $W/\gamma=0.8$ in the phase encoding of the classical control in the target states. The last panel displays the absolute difference between the target and reconstructed Wigner functions.
			(c) The valid prediction time in classical (C-VPT) and quantum tomography (Q-VPT) tasks depending on the input scaling $W/\gamma$ and the frequency of the sinusoidal classical inputs.
			\label{fig:close:phase}}
	\end{figure}

	Figure~\ref{fig:close:amp}(a) (for the amplitude encoding of the classical control in the target state) and 
	Fig.~\ref{fig:close:phase}(a) (for the phase encoding of the classical control in the target state) 
	present examples of the classical control signals in the training and closed-loop phase for $f=60$ with different input scaling $W$.
	Since we consider the same input sequence for both encoding methods, the results for this classical prediction are the same.
	For $W/\gamma=0.8$, the control signal is almost reconstructed perfectly for all time steps in the closed-loop phase with a QR of $N=3$ sites and a measurement multiplexity $V=10$.
	Subsequently, as shown in Fig.~\ref{fig:close:amp}(b), this QR can efficiently reconstruct the Wigner function even without accessing the control signal in the closed-loop phase since the errors in each coordinate $(x_j, p_j)$ are almost zero.
	The errors for the tomography task in Fig.~\ref{fig:close:phase}(b) are larger but still less than 0.05, demonstrating that the phase encoding of classical input in the output $\cF_l$ is more difficult to emulate than the amplitude encoding. 
	
	We evaluate the performance of the closed-loop control using the valid prediction time (VPT) in the classical and quantum tomography tasks.
	Assume that the training phase is performed for the index $l=1,\ldots,L$. Given a positive integer $t$, we define the following errors $\textup{NRMSE}(t)$ and $\textup{EW}(t)$, which are the prediction errors of the classical and quantum tomography tasks in the closed-loop phase, respectively.
	
	\begin{align}
		\textup{NRMSE}(t) &= \sqrt{\dfrac{1}{t}\sum_{l=L}^{L+t-1}\dfrac{(s_{l+1} - v_{l})^2}{\sigma^2_s} },\\
		\textup{EW}(t) &= \sqrt{ \dfrac{1}{t}\sum_{l=L+1}^{L+t}\dfrac{\sum_{i,j}[\cW(\cF_l; x_i, p_j) - \hat{\cW}(\cF_l; x_i, p_j)]^2}{\sum_{i,j}[\cW(\cF_l; x_i, p_j) + \hat{\cW}(\cF_l; x_i, p_j)]^2}},
	\end{align}
	where $\sigma^2_s$ is the variance of the classical control signal $\{s_l\}$.    
	Given an error threshold $\varepsilon$, the valid prediction time is defined as the longest time for which the error is smaller than or equal to $\varepsilon$:
	
	\begin{align}
		\textup{C-VPT}(\varepsilon) &= \max \{ T \mid  \textup{NRMSE}(t) \leq  \varepsilon \quad \forall t \leq T\} \\
		\textup{Q-VPT}(\varepsilon) &= \max \{ T \mid  \textup{EW}(t) \leq  \varepsilon \quad \forall t \leq T\}.
	\end{align}

	We plot the dependency of the VPTs on the scaling $W$ and frequency $f$ of the control signals [Fig.~\ref{fig:close:amp}(c) and Fig.~\ref{fig:close:phase}(c)].
	For the amplitude encoding of classical control in the target states [Fig.~\ref{fig:close:amp}(c)], we observe an optimal range ($0.8 \leq W/\gamma \leq 1.2$) of input scaling $W$ for optimal performance of the tomography task.
	Here, we note that setting a too large $W$ will impose the localization in the quantum dots and may lead to non-ergodic behavior in the QR.
	In this case, when the input state $\beta_l$ is incident to the QR with weak coupling ($|W_{jk}^{\textup{in}}| \ll |P + Ws_l|$), sufficient information regarding $\beta_l$ cannot be extracted from the QR.
	For the phase encoding of the classical control in the target states [Fig.~\ref{fig:close:phase}(c)], we observe a trade off in the values of C-VPT and Q-VPT, where setting a small $W$ ($0.1 \leq W/\gamma  \leq 0.8$) can increase the C-VPT but decrease the Q-VPT.
	The results in Fig.~\ref{fig:close:amp}(c) and Fig.~\ref{fig:close:phase}(c) show that different temporal quantum maps $\cF$ require different profiles of information processing, which can be adjusted by some modifiable parameters, such as the input scaling $W$ or the constant coherent field $P$. 
	It can be helpful if we can evaluate the required information processing ability of a temporal quantum map, such as how far and which combinations of the previous inputs are processed in this map. This question is directly related to the  information processing framework in input-driven dynamical systems~\cite{dambre:2012:nonlinear,kubota:2021:prr:IPC} but presents further challenges in our hybrid setting. 
	
	\begin{figure}
		\includegraphics[width=16cm]{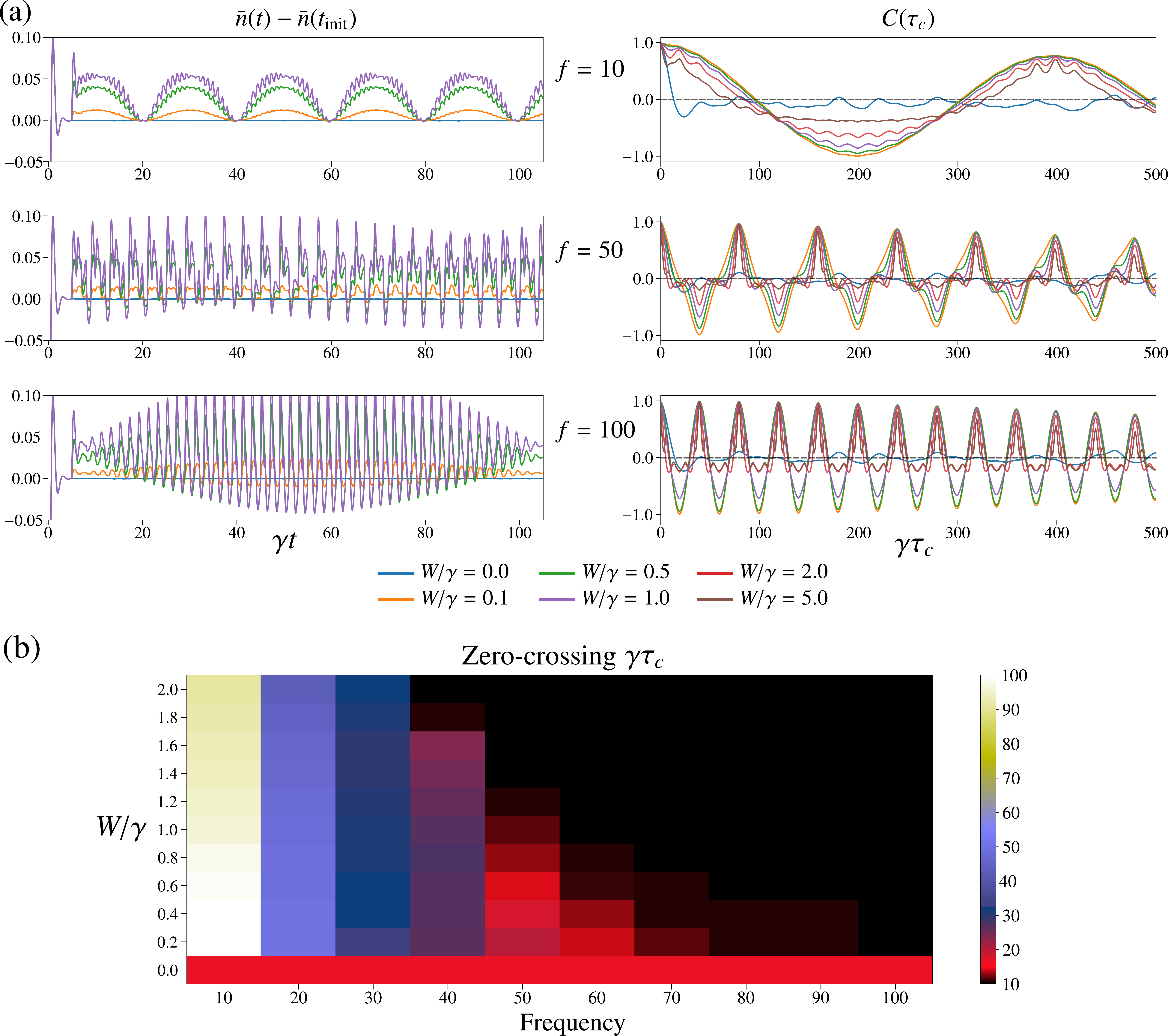}
		\protect\caption{(a) The dynamics of the average occupation numbers $\bar{n}(t)$ over $N=3$ reservoir sites (left panel) and the autocorrelation for one trial of the QR's configuration and data (right panel) at different $f$ and input scaling $W/\gamma$. 
			(b) The average values of autocorrelation zero-crossing times over 10 trials of the QR's configurations and input data for each combination of $W/\gamma \in \{0.0, 0.2, \ldots, 1.8, 2.0\}$ and $f \in \{10, 20, \ldots, 90, 100\}$.
			\label{fig:cross:delay}}
	\end{figure}
	
	We also notice the dependency of the VPTs on the time scales of the classical control signals in Fig.~\ref{fig:close:amp}(c) and Fig.~\ref{fig:close:phase}(c), where the classical inputs with higher frequencies $f$ (lower timescales) basically lead to better performance.
	To characterize the activity of the QR, we observe the dynamics of the average occupation numbers $\bar{n}(t)$ over reservoir sites at different $f$ and input scaling $W$ [left panels in Fig.~\ref{fig:cross:delay}(a)]. 
	In the presence of periodic classical inputs ($W/\gamma > 0.0$), an oscillatory response is superposed on the intrinsic dynamics of the quantum input without the classical input ($W/\gamma=0.0$).
	We further calculate the autocorrelation function of each frequency $f$ averaged across all the reservoir sites,
	\begin{align}
		C(\tau_c) = \dfrac{1}{N}\sum_{j=1}^N \langle \left( n_j(t) - \langle  n_j(t) \rangle  \right)\left( n_j(t+\tau_c) - \langle  n_j(t) \rangle  \right) \rangle,
	\end{align}
	where the angular brackets denote a time average.
	Here, $C(0)$ depicts the total variance in the fluctuations of the occupation numbers in the reservoir sites, whereas $C(\tau_c)$ with $\tau_c > 0$ provides information about the temporal structure of the reservoir activity.
	In the right panels of Fig.~\ref{fig:cross:delay}(a), we plot the autocorrelation for one trial of the QR's configuration and data at different $f$ and input scaling $W$.
	With no classical input ($W/\gamma = 0.0$), the autocorrelation function decays to the values around zero as $\tau_c$ increases. This implies that temporal fluctuations are uncorrelated over large time intervals, which is due to the effect of random quantum inputs and disordered dynamics in the QR.
	When the QR is driven by sinusoidal classical inputs, we observe that the periodic activity induced by these inputs is superposed on the background of the quantum inputs.
	
	We define the timescale of the QR as the first $\tau_c$ such that $C(\tau_c)$ crosses the zero line, which can be understood as the first time interval where the temporal temporal fluctuations are uncorrelated.
	This zero-crossing time depends on the spontaneous activity of the QR and the timescale of the external classical input.
	We plot in Fig.~\ref{fig:cross:delay}(b) the average values of zero-crossing times over 10 trials of the QR's configurations and input data for each combination of $W/\gamma \in \{0.0, 0.2, \ldots, 1.8, 2.0\}$ and $f \in \{10, 20, \ldots, 90, 100\}$.
	In the presence of external classical inputs ($W/\gamma > 0.0$), if the zero-crossing times are larger than those of no classical inputs ($W/\gamma = 0.0$), we observe low values of VPTs in Fig.~\ref{fig:close:amp}(c) and Fig.~\ref{fig:close:phase}(c).
	These results imply that the timescales of the QRs without classical inputs should be larger than the timescales induced by classical inputs. These timescales can be modified by adjusting the constant coherent field $P$ and the input scaling $W$.

\begin{figure}
		\includegraphics[width=18cm]{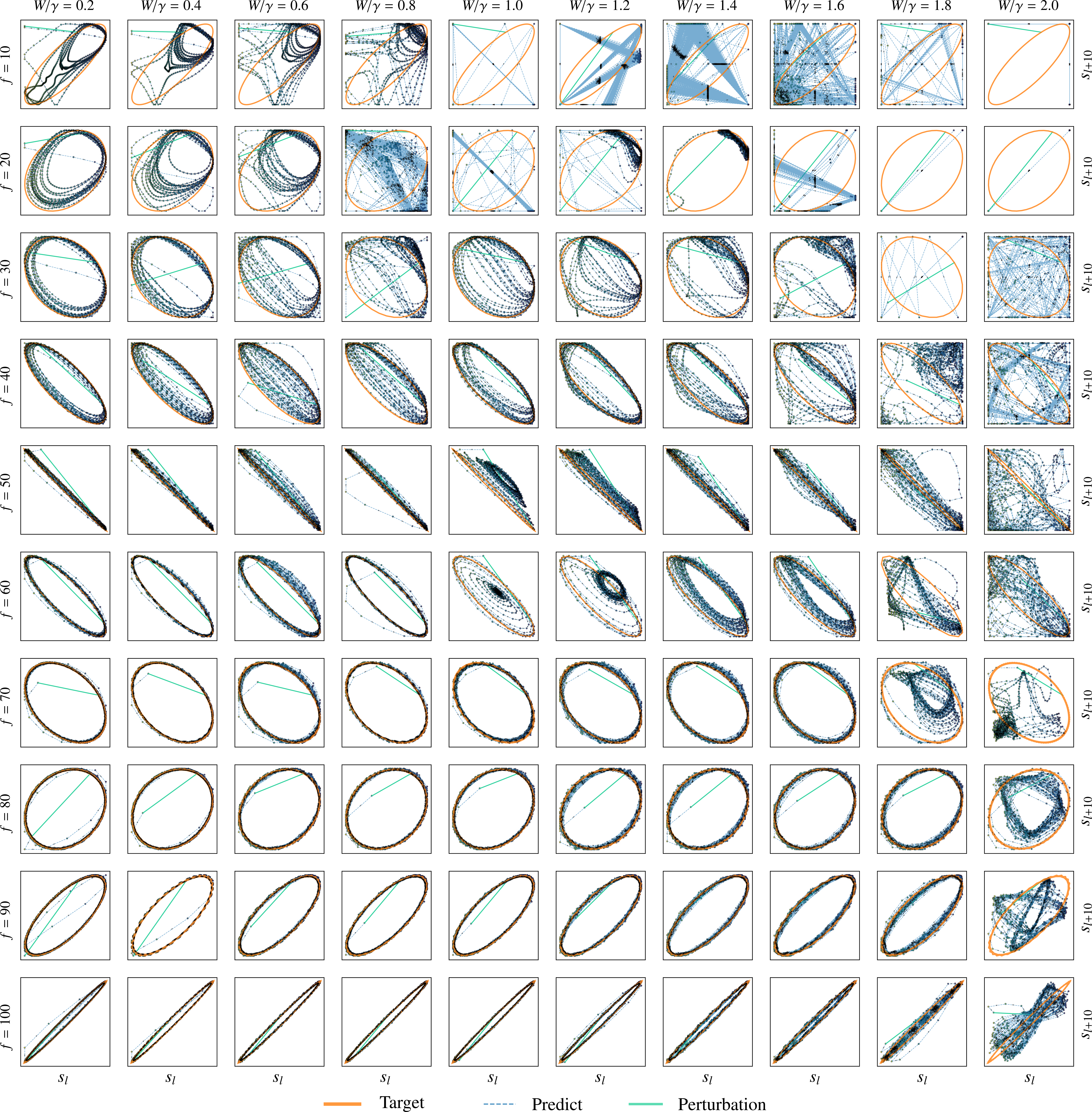}
		\protect\caption{Stability after adding a small perturbation to the trajectory. The QR presents a stable embedding of sinusoidal classical inputs if the trajectory can return to the target after adding a small perturbation (green line) into the predicted value. We observe appropriate ranges of input scaling $W$ and $f$ to obtain stable closed loops.
			There is an intriguing observation that if we increase the input scaling $W/\gamma$, the closed loop fails to reconstruct the trajectory of sinusoidal inputs but can produce chaotic-like behavior in the embedding space.
			\label{fig:close:pert}}
	\end{figure}

\begin{figure}
		\includegraphics[width=18cm]{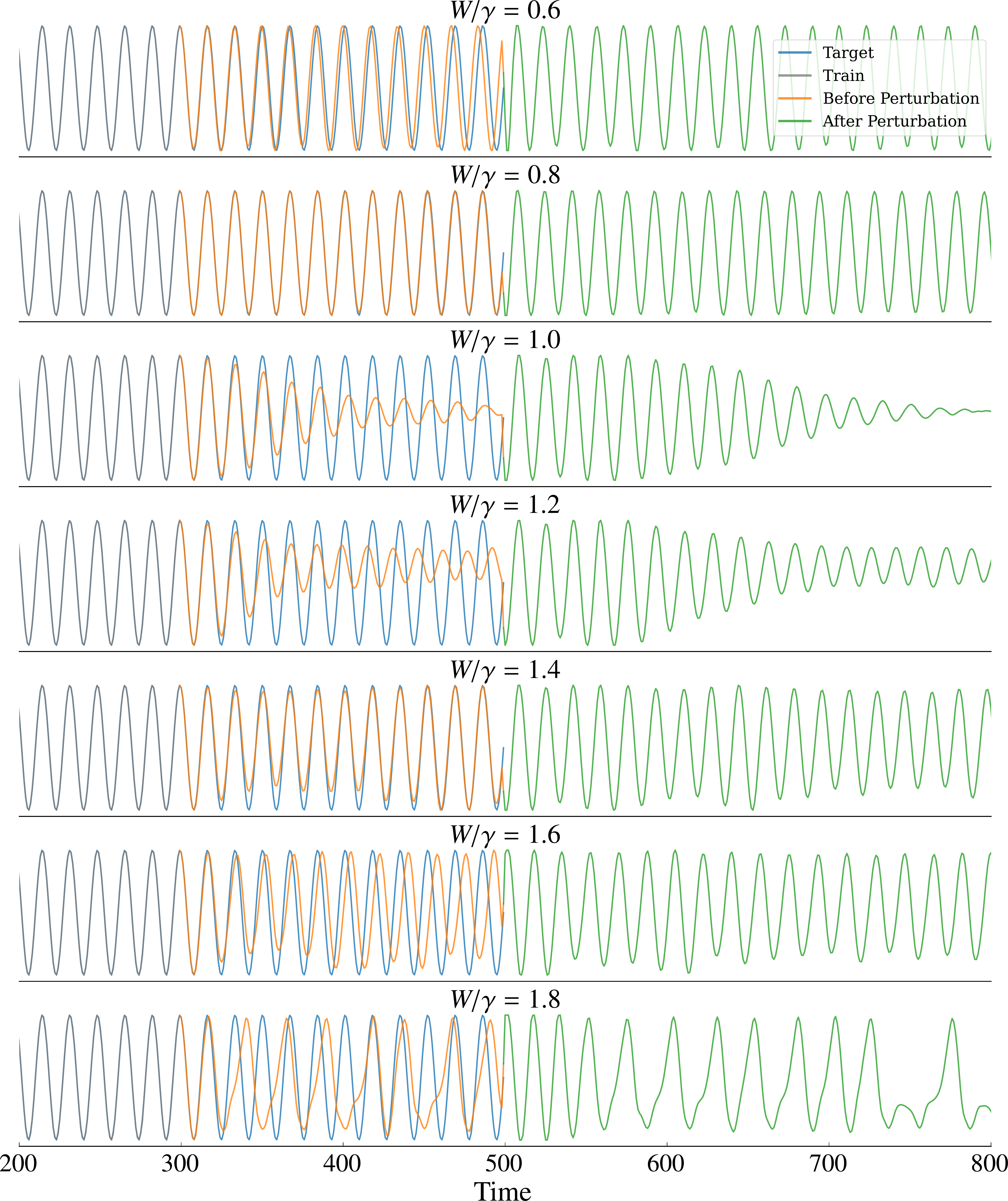}
		\protect\caption{Stability after adding a small perturbation to the sinusoidal classical input with $f=60$ and different values of $W/\gamma$. The QR presents a stable reconstruction if the trajectory can return to the target after adding a small perturbation into the predicted value. In this experiment, we observe that $W/\gamma=0.8$ can provide a stable reconstruction.
			\label{fig:series:pert}}
	\end{figure}

	We further analyze the effect of perturbation to investigate the stability of the embedded classical trajectories. Figure~\ref{fig:close:pert} shows the output dynamics of both the target and perturbed prediction trajectories in the closed-loop phase plotted in the $(s_l, s_{l+10})$ plane for different values of $f$ and $W/\gamma$. 
	After the training phase, we add a small perturbation into the predicted value,
	which results in an extra drift in the $(s_l, s_{l+10})$ plane (green line).
	The reservoir presents a stable embedding of sinusoidal classical inputs if the trajectory can return to the target one after the addition of the perturbation.
	We observe appropriate ranges of input scaling $W/\gamma$ and $f$ to obtain stable closed loops (Fig.~\ref{fig:close:pert} and Fig.~\ref{fig:series:pert}).
	Furthermore, if we increase the input scaling $W/\gamma$, the closed loop fails to reconstruct the trajectory of sinusoidal inputs but can produce chaotic-like behavior in the embedding space.
	Intriguingly, the generated trajectory is not elliptical as the trajectory of sinusoidal inputs but is still robust with respect to a small perturbation.
	
	\clearpage
	\section{Quantum Memory Capacity Defined via Tomography Learning}
	In Fig.~2 in our main text, we present the results of tomography for the quantum switch, which requires the information of previous input signals.
	The performance of this task depends on the amount of memory from previous inputs that the classical readout can retrieve from the reservoir.
	
	In the conventional RC, we evaluate the short-term memory (STM) property of the reservoir via the delay-reconstruction task for reconstructing the previous input. Given a time delay $d \geq 0$ and a uniform random input sequence $\{u_n\}$, the target of this task is to produce the output sequence $\{y_n\}$ such as $\{y_n\}$ can approximate the target sequence $\{\hat{y}_n = u_{n-d}\}$.
	For each delay time step $d$, the readout part is trained to remember the input sequences at delayed $d$-time steps.
	The performance is evaluated by the square of the correlation coefficient $\cC(d)$ between the output and delayed input sequences~\cite{jaeger:2001:short} as follows:
	\begin{align}
		\cC^2(d) = \dfrac{{\textup{cov}}^2(\{y_n\}, \{u_{n-d}\} )}{\textup{var}(\{y_n\})\textup{var}(\{u_n\})}.
	\end{align}
	Here, $\textup{cov}(\cdot)$ and $\textup{var}(\cdot)$ denote the covariance and variance function, respectively.
	The STM property represents that this $\cC^2(d)$ is sufficiently small at large values of the delay $d$.
	$\cC^2(d)$ is defined as the \textit{memory function} to characterize the memory profile of the reservoir.
	Furthermore, the \textit{memory capacity} (MC) of the reservoir is given by
	\begin{align}
		\textup{MC} = \sum_{d=0}^{\infty}\cC^2(d).
	\end{align}
	
	In our study, we consider the concept of quantum memory capacity~\cite{tran:2021:temporal} to measure the ability of the QR to reconstruct the previous quantum inputs via the classical readout.
	We investigate the quantum version of STM in the QR via the quantum version of the delay-reconstruction task $\cF(u_n=0, \beta_n) = \beta_{n-d}$ given the delay $d$, where classical inputs are zero.
	Since the input and output are quantum states, the capacity to reconstruct the previous $d$ steps of the input states is evaluated via the square of the distance correlation~\cite{gabor:2007:corr} between the output $\{\sigma_n\}$ (obtained via the training of the classical readout) and the target $\{\hat{\sigma}_n\}=\{\beta_{n-d}\}$:
	\begin{align}
		\cR^2(d) = \dfrac{\cV^2(\{\sigma_n\}, \{\beta_{n-d}\})}{\sqrt{\cV^2(\{\sigma_n\}, \{\sigma_n\})\cV^2(\{\beta_{n}\}, \{\beta_{n}\})}}.
	\end{align}
	Here, $\cV^2(\{\rho_n\}, \{\sigma_n\})$ represents the squared distance covariance of random sequences of density matrices $\{\rho_n\}, \{\sigma_n\}$.
	The squared distance covariance $\cV^2(\{\rho_n\}, \{\sigma_n\})$ is calculated from all pairwise distances $A(\rho_j, \rho_k)$ and $A(\sigma_j, \sigma_k)$ for $j, k = 1,2,\ldots,n$,
	where the distance $A(\rho, \sigma) = \arccos{F(\rho, \sigma)}$ for given density matrices $\rho$ and $\sigma$ is defined as the angle induced from the fidelity $F(\rho, \sigma)=\tr[\sqrt{\sqrt{\sigma}\rho\sqrt{\sigma}}]$.
	We construct the distance matrices for $\{\rho_n\}$ and $\{\sigma_n\}$ as $(R_{jk})$ and $(S_{jk})$ with the elements $R_{jk}=A(\rho_j, \rho_k)$ and $S_{jk}=A(\sigma_j, \sigma_k)$.
	We take all double centered distances
	\begin{align}
		r_{j,k} &= R_{j,k} - \bar{R}_{j.} - \bar{R}_{.k} + \bar{R}_{..},\\
		s_{j,k} &= S_{j,k} - \bar{S}_{j.} - \bar{S}_{.k} + \bar{S}_{..},
	\end{align}
	where $\bar{R}_{j.}$ and $\bar{R}_{.k}$ are the $j$th row mean and the $k$th column mean, respectively,
	and $\bar{R}_{..}$ is the grand mean of the distance matrix $(R_{jk})$ (the same notations for $S$).
	The squared distance covariance is the arithmetic average
	\begin{align}
		\cV^2(\{\rho_n\}, \{\sigma_n\}) = \dfrac{1}{n^2}\sum_{j=1}^n\sum_{k=1}^nr_{j,k}s_{j,k}.
	\end{align}
	
	$\cR^2(d)$ gives information about the serial dependence between $\{\sigma_n\}$ and $\{\hat{\sigma}_n\}=\{\beta_{n-d}\}$.
	Here, $0\leq \cR^2(d) \leq 1$ and $\cR^2(d) = 1$ if we can find some linear transformation from the output sequence $\{\sigma_n\}$ to the target sequence $\{\hat{\sigma}_n\}$.
	In contrast, $\cR^2(d)=0$ implies that the system cannot reconstruct the previous $d$ steps of the inputs because the output and the target sequences are completely independent.
	We define $\cR^2(d)$ as the \textit{quantum memory function} of the QR via tomography learning with the classical readout.
	Consequently, the \textit{quantum memory capacity} (QMC) is defined as  
	\begin{align}\label{eqn:qmc}
		\textup{QMC} = \sum_{d=0}^{\infty}\cR^2(d).
	\end{align}
	
\begin{figure}
		\includegraphics[width=17cm]{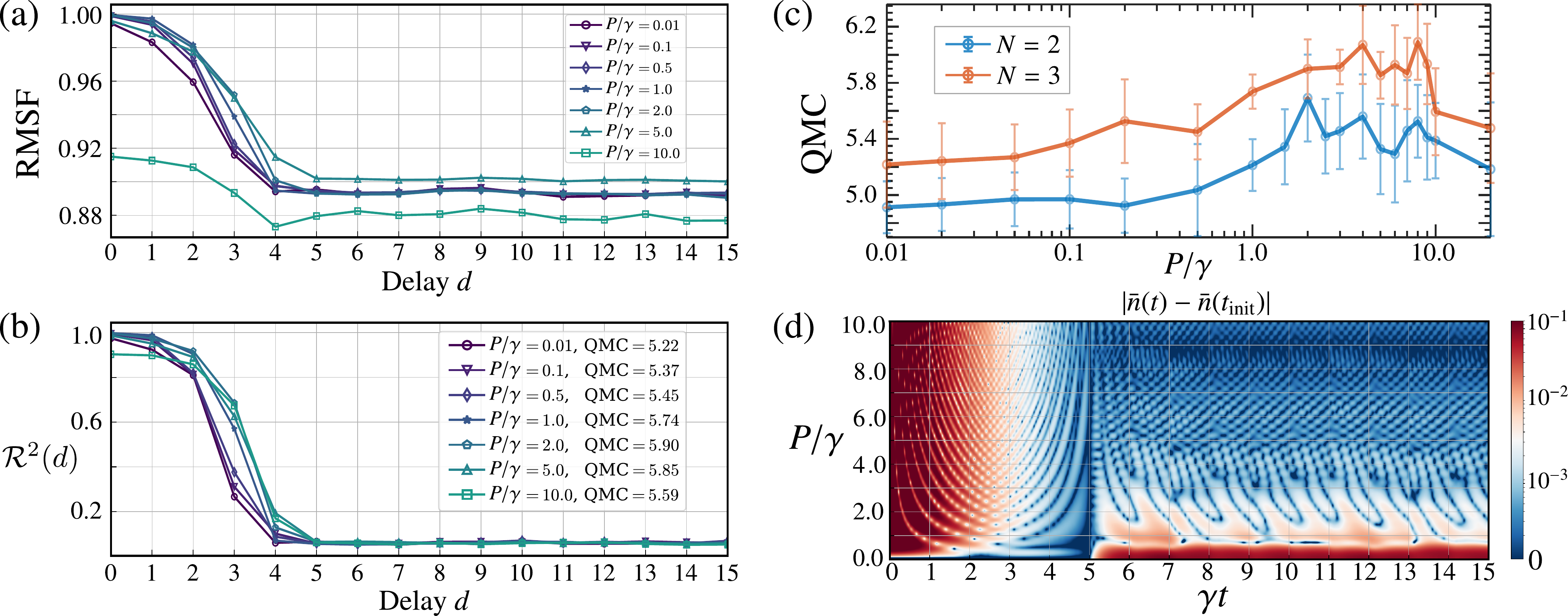}
		\protect\caption{(a) The RMSF of the STM task varying by the delay $d$ and (b) the quantum memory function $\cR^2(d)$ for $N=3$-site QR at several values of uniform excitation $P(t)=P$.
			The tomography task is performed with $V=5$ measurement multiplexity and a random sequence of one-qubit input states with 400 time steps for the training and 100 time steps for the evaluation.
			(c) The dependency of quantum memory capacity (QMC) on the coherent field strength $P$ at $N=2,3$-site QR. The QMC is averaged over 10 random trials with displayed error bars.
			(d) The absolute difference $| \bar{n}(t) - \bar{n}(t_{\textup{init}}) |$ between the average occupation numbers $\bar{n}(t)$ at initial time $t_{\textup{init}}=5/\gamma$ and an arbitrary $t$ $(0\leq \gamma t \leq 15)$ varying by $P/\gamma$. The quantum input is incident on the QR at $\gamma t=5, 6, 7, \ldots, 14, 15$, which increases the value $| \bar{n}(t) - \bar{n}(t_{\textup{init}}) |$ before decreasing it until the next input.
			\label{fig:qmem}}
	\end{figure}

	Figure~\ref{fig:qmem} presents a demonstration for the quantum memory function and quantum memory capacity, where we consider $\{\beta_n\}$ as a random sequence of one-qubit input states with 400 time steps for the training and 100 time steps for the evaluation.
	Figure~\ref{fig:qmem}(a) displays the values of RMSF, and Fig.~\ref{fig:qmem}(b) displays the quantum memory function $\cR^2(d)$ for $N=3$-site QR at several values of uniform excitation $P(t)=P$.
	We perform the tomography task with $V=5$ measurement multiplexity, which means that the dimension of the reservoir state is $VN=15$.
	The STM property depends on the value of $P$, where the memories at $d < 5$ dominate all the regions and converge to almost the same value at a sufficiently large $d$. This value is non-zero owing to the effect of the finite data length.
	
	We further plot the dependency of quantum memory capacity (QMC) on the coherent field strength $P$ in Fig.~\ref{fig:qmem}(c) at $N=2,3$-site QR.
	Here, Eq.~\eqref{eqn:qmc} is calculated until the maximum delay $d_{\textup{max}}=40$.
	We observe an optimal region of $P$ ($2\leq P < 10$), where the QMC is favorable. 
	To explain this behavior, we further analyze the dynamics of the occupation numbers $n_j(t)$ in the reservoir sites.
	Figure~\ref{fig:qmem}(d) plots the absolute difference $| \bar{n}(t) - \bar{n}(t_{\textup{init}}) |$ between the average element $\bar{n}(t)$ of the reservoir states at initial time $t_{\textup{init}}=5/\gamma$ and an arbitrary $t$ $(0\leq \gamma t \leq 15)$.
	This difference approaches zero as $t$ approaches $t_{\textup{init}}$.
	The quantum input is incident to the QR at $\gamma t=5, 6, 7, \ldots, 14, 15$, which increases the value $| \bar{n}(t) - \bar{n}(t_{\textup{init}}) |$ before decreasing it until the next input.
	We anticipate that increasing the magnitude of the coherent field strength $P$ compared with $h_{ij}$ in Eq.~\eqref{eqn:hamiltonian} may lead to non-ergodic behavior in the QR, i.e., a strong and qualitative dependence of expectation values on the initial state at $t_{\textup{init}}$ (Fig.~\ref{fig:tau}).
	Furthermore, the input state $\beta_l$ is incident to the QR with weak coupling ($|W_{jk}^{\textup{in}}| \ll |P|$) in this case.
	Therefore, sufficient information regarding $\beta_l$ cannot be extracted from the QR as $| \bar{n}(t) - \bar{n}(t_{\textup{init}}) | \approx 0$. 
	In contrast, a small $P(t)$ strongly drives the system from the steady state at the input-injecting timing but reduces the memory effect of the QR in reconstructing past information since the old information is replaced very quickly.
	
	\clearpage
	\section{Effects of Classical Input on the Reconstruction of Quantum Input}
	
	In the main text, we considered the target function, which is a function $\cF$ of hybrid inputs $(u, \beta)$.
	In this case, information of both the classical input $u(t)$ and quantum input $\beta(t)$ is retained in the reservoir states.
	Since the classical input $u(t)$ is encoded into the strength of the coherent field $P(t) = P + Wu(t)$, the classical input and input scaling $W/\gamma$ have a strong effect on the dynamics of the QR.
	If the target function $\cF$ does not depend on the classical input $u(t)$, then the injection of $u(t)$ into the QR may affect the reconstruction of $\cF$.
	
	In this section, we verify this observation by investigating the memory capacity.
	Given a sequence of hybrid inputs $\{u_n, \beta_n\}$, we use our QR in a multitask setting with the classical and quantum delay-reconstruction tasks mentioned in the previous section.
	Given a delay $d$, we consider the delay reconstruction of the classical input $\{u_{n-d}\}$ in the classical task and the delay reconstruction of the quantum input $\{\beta_{n-d}\}$ in the quantum task.
	We compute the corresponding MC and QMC for the classical and quantum tasks, respectively.
	
\begin{figure}
		\includegraphics[width=16cm]{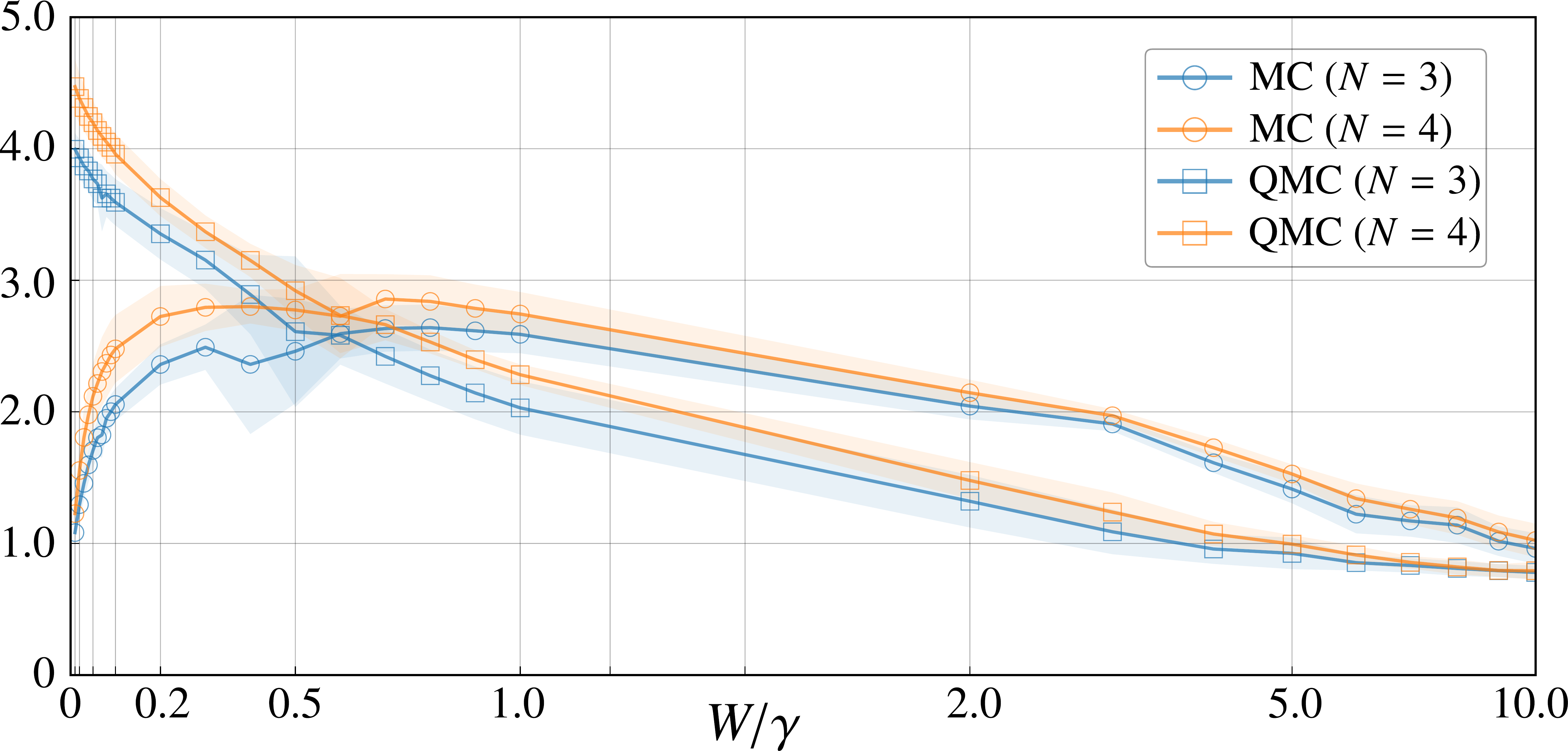}
		\protect\caption{The dependency of MC and QMC on the input scaling $W/\gamma$ for $N=3,4$-site QR with a constant coherent field strength $P/\gamma=1.0$. We consider the random uniform $\{u_n\}$ for classical inputs and $\{\beta_n\}$ as a random sequence of one-qubit states for quantum inputs with 800 time steps for the training and 200 time steps for the evaluation.
			The tomography task is performed with $V=8$ measurement multiplexity, and MC and QMC are calculated until the maximum delay $d_{\textup{max}}=20$. 
			The solid lines and the shaded areas indicate the median values and the confidence intervals (one standard deviation) calculated in the ensemble of 10 random trials of the input sequence and the QR's configuration, respectively.
			\label{fig:mem:tradeoff:cv0}}
	\end{figure}

	Figure~\ref{fig:mem:tradeoff:cv0} displays the dependency of MC and QMC on the input scaling $W/\gamma$ for $N=3,4$-site QR with a constant coherent field strength $P/\gamma=1.0$. Here, we consider the random uniform $\{u_n\}$ for classical inputs and $\{\beta_n\}$ as a random sequence of one-qubit states for quantum inputs.
	To attain the same setting that was used in the task described in Fig.~2 in the main text, we perform the tomography with $V=8$ measurement multiplexity and use 800 time steps for the training and 200 time steps for the evaluation. We compute the MC and QMC until the maximum delay $d_{\textup{max}}=20$. 
	The result demonstrates that the QMC is reduced when the random classical input is introduced into the QR with  increasing input scaling $W/\gamma > 0$.
	For a relatively large $W/\gamma > 1.0$, both MC and QMC decrease owing to the localization effect with a large strength of the coherent field $P(t)=P + Wu(t)$
	However, at $W/\gamma \leq 1.0$, we observe a trade-off relation between MC and QMC. Here, increasing $W/\gamma$ from zero can help improve the MC but reduce the QMC.
	This observation implies that this QR may not perform well if the target function does not depend on the classical input, and the fluctuation of classical inputs has a strong effect on the QR dynamics. 
	However, if the target function is the function of the classical and quantum input, we can use the trade-off of MC and QMC to adjust $W/\gamma$ for an optimal performance.
	
	\clearpage
	\section{Quantum Readout for Temporal Quantum Learning}
	
	\begin{figure}
		\includegraphics[width=8.5cm]{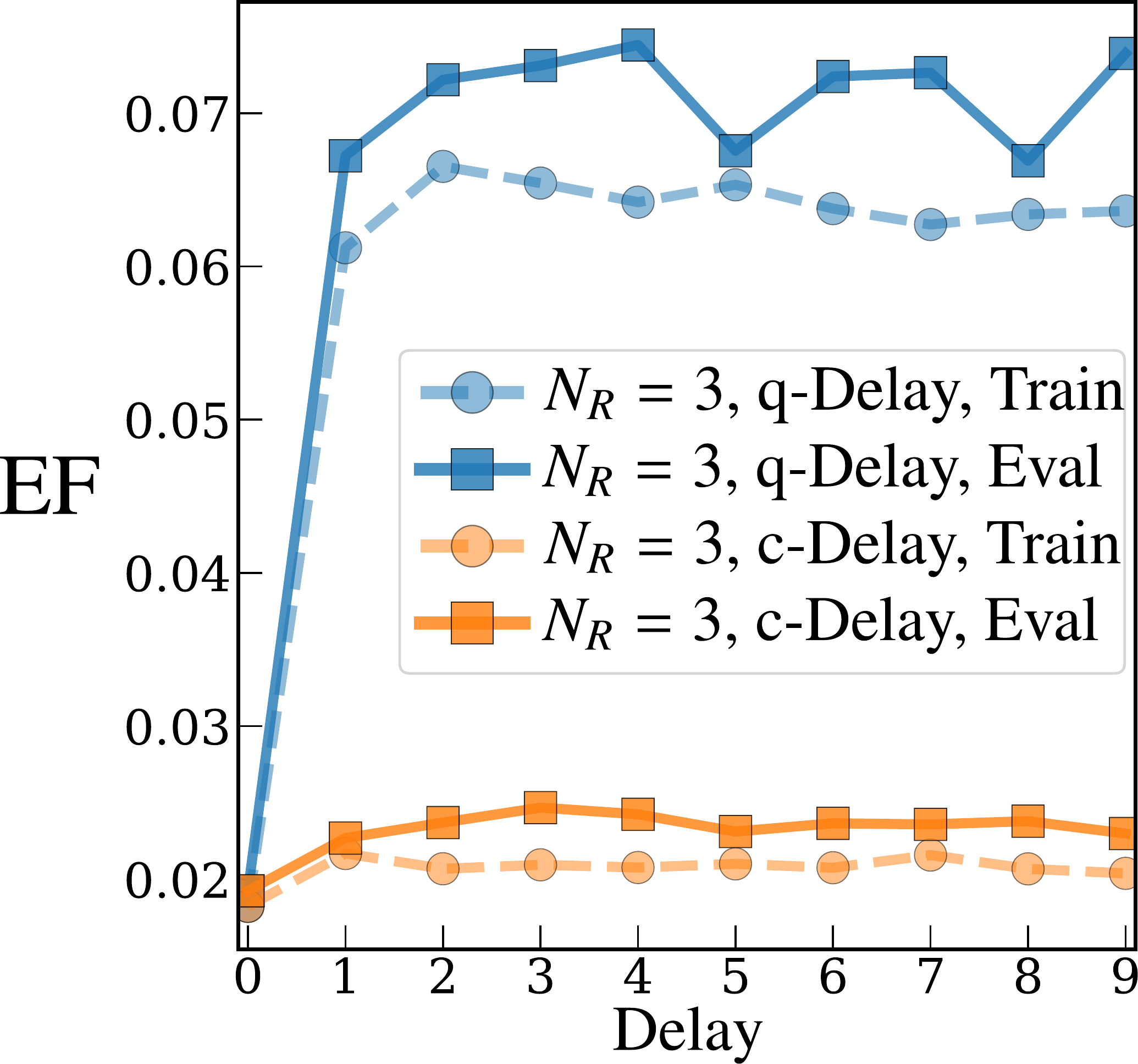}
		\protect\caption{The error evaluated to prepare the temporal depolarizing quantum channel for varying the delays. 
			We consider both input modes $\ha_k$ and reservoir modes $\hc_i$ as $N_R=3$ readout nodes and train both interaction coefficients $W^{\textup{in}}_{jk}$ and readout weights.
			The blue lines represent $d_c=0$ and the variation of $d_q$. The orange lines represent $d_q=0$ and the variation of $d_c$.
			\label{fig:qtask:depolarheat}}
	\end{figure}
	
	We present a proof of concept application for quantum tasks using the quantum readout scheme. We use the QR to prepare the target quantum state of a temporal depolarizing quantum channel $\cF\{(s_l, \beta_l)\} = s_{l-d_c}I/D + (1-s_{l-d_c})\beta_{l-d_q}$, where $\{\beta_l\}$ is randomly generated in a $D$-dimensional Hilbert space, $\{s_l\}$ is a random sequence in $[0, 1]$, and $d_c, d_q$ are the delay times.
	Here, we consider a sequence of $200$ one-qubit quantum states for the training and $100$ states for the evaluation. The baseline is computed when we set the output the same as the input.
	We use the Nelder\text{--}Mead simplex algorithm~\cite{lagarias:1998:nd} to minimize 
	the fidelity error
	$
	\textup{EF}=\sqrt{(1/L)\sum_{l=1}^{L}[1-F(\sigma_l, \hat{\sigma}_l)]^2},
	$
	where $\sigma_l$ and $\hat{\sigma}_l$ are the target and the preparing quantum states, respectively.
	
	Figure~\ref{fig:qtask:depolarheat} illustrates the average fidelity error as a function of 
	$d_c$ and $d_q$ over 10 trials. Here, we consider both input modes $\ha_k$ and reservoir modes $\hc_i$ as $N_R=3$ readout nodes, and both interaction coefficients $W^{\textup{in}}_{jk}$ and readout weights as the training parameters. For $d_q=0$ (orange lines), we can prepare the target channel with a small $d_c < 3$. However, if $d_q > 0$ (blue lines with $d_c=0$), the large increases in the fidelity errors imply that it is difficult to realize previous quantum inputs, which may incur a higher cost for training more readout nodes.

\begin{figure}
		\includegraphics[width=15cm]{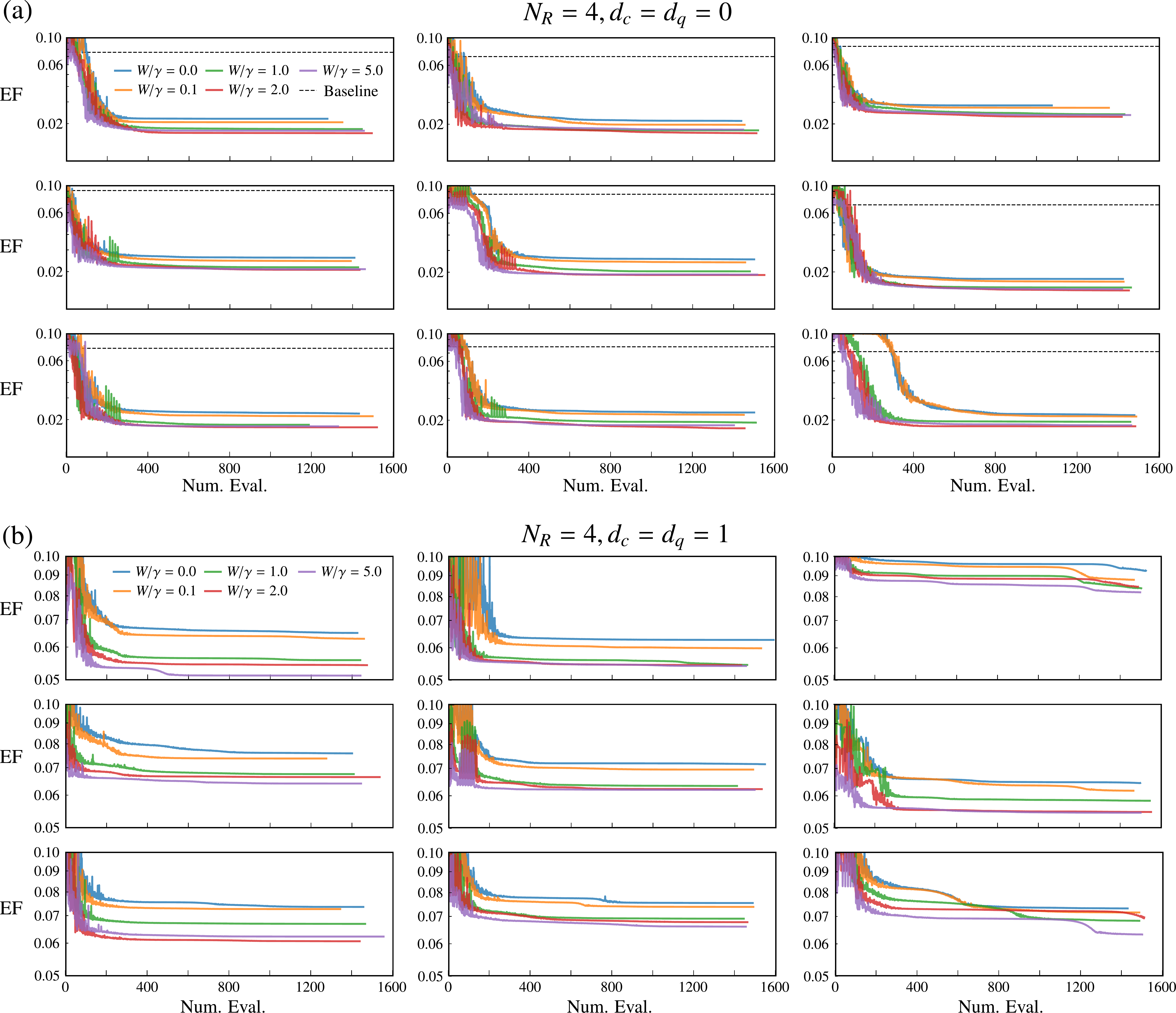}
		\protect\caption{Nelder\text{--}Mead steps for minimizing the fidelity errors EF with (a) ($d_c=d_q=0$) and (b) ($d_c=d_q=1$) at different values of the input scaling $W/\gamma$. Each graph represents each trial.
			Here, we consider both input modes $\ha_k$ and reservoir modes $\hc_i$ as $N_R=4$ readout nodes and train both in- and out-weight parameters.
			\label{fig:train:delay}}
	\end{figure}

	To understand the training process, we show the Nelder\text{--}Mead steps for minimizing the fidelity errors EF in Fig.~\ref{fig:train:delay}(a) ($d_c=d_q=0$) and Fig.~\ref{fig:train:delay}(b) ($d_c=d_q=1$) at different values of the input scaling $W$. 
	Here, we consider $N_R=4$ readout nodes and 9 trials of input data and the QR's configurations. 
	The process starts with an initial guess for which EF is large and progressively reaches to a minimum value.
	The training results are still below the considerable good preparations, especially with a non-zero delay. However, they demonstrate that the hybrid quantum-classical inputs are effectively considered in training the quantum readout.
	
	\clearpage
	
\providecommand{\noopsort}[1]{}\providecommand{\singleletter}[1]{#1}%